\documentclass[a4paper,onecolumn,11pt]{quantumarticle}
\pdfoutput=1

\usepackage[english]{babel}
\usepackage[T1]{fontenc}

\DeclareUnicodeCharacter{2082}{$_2$}
\usepackage{algorithm}
\usepackage{algpseudocode}
\usepackage{setspace}
\usepackage{enumitem}
\usepackage{authblk}
\usepackage{colortbl}
\usepackage{amsmath, amssymb, amsfonts}
\usepackage{braket}
\usepackage{physics} 
\newcommand{\braopket}[3]{\langle #1 | #2 | #3 \rangle}
\usepackage{caption}
\usepackage{color}
\usepackage{graphicx}
\usepackage{tikz}

\usepackage{booktabs}
\usepackage{longtable}
\usepackage{adjustbox}
\usepackage{lscape}
\usepackage{xcolor}
\usepackage{multirow}

\usepackage[numbers,sort&compress]{natbib}

\usepackage{hyperref}
\hypersetup{
    bookmarksnumbered=true, 
    bookmarksdepth=-2
}

\graphicspath{{./}} 

\begin{document}

\title {Compactifying Electronic Wavefunctions I: Error-Mitigated Transcorrelated DMRG}

\newcommand{\cquic}{Center for Quantum Information and Control, Department of Physics and Astronomy, University of New Mexico, Albuquerque, New Mexico 87131, USA}
\newcommand{\ufpe}{Departamento de Física - Universidade Federal de Pernambuco, Av. Prof. Moraes Rego, 1235, 50670-901 Recife, PE, Brazil}

\author[1,2]{Bruna G. M. Ara{\'{u}}jo}\thanks{bruna.gabriellyma@gmail.com}
\author[2]{Antonio M. S. Mac\^edo}\thanks{antonio.smacedo@ufpe.br}

\affil[1]{\cquic}
\affil[2]{\ufpe}

\orcid{0000-0001-9729-6001}

\begin{abstract}
Transcorrelation (TC) techniques effectively enhance convergence rates in strongly correlated fermionic systems by embedding electron-electron cusp into the Jastrow factor of similarity transformations, yielding a non-Hermitian, yet iso-spectral, Hamiltonian. This non-Hermitian nature introduces significant challenges for variational methods such as the Density Matrix Renormalization Group (DMRG). To address these, existing approaches often rely on computationally expensive methods prone to errors, such as imaginary-time evolution. We introduce an Error-Mitigated Transcorrelated DMRG (EMTC-DMRG), a classical variational algorithm that overcomes these challenges by integrating existing techniques to achieve superior accuracy and efficiency. Key features of our algorithm include: (a) an analytical formulation of the transcorrelated Fermi-Hubbard Hamiltonian; (b) a numerically exact, uncompressed Matrix Product Operator (MPO) representation developed via symbolic optimization and the Hopcroft-Karp algorithm; and (c) a time-independent DMRG with a two-site sweep algorithm; (d) we use Davidson solver even for a non-Hermitian Hamiltonian. Our method significantly enhances computational efficiency and accuracy in determining ground-state energies for the two-dimensional transcorrelated Fermi-Hubbard model with periodic boundary conditions. Additionally, it can be adapted to compute both ground and excited states in molecular systems.
\end{abstract}
\tableofcontents 
\section{Introduction}

Efficiently addressing electronic Hamiltonians is paramount in research areas such as quantum chemistry, materials science, and condensed matter physics. A range of approaches have been employed to achieve this goal, such as full configuration interaction quantum Monte Carlo (FCIQMC) \cite{ref1_intro, ref2_intro} and various Density Matrix Renormalization Group (DMRG) formulations \cite{ref8_22_intro}. Given the inherent complexity of these systems, understanding the nature of electron correlation is essential, as it dictates the accuracy and efficiency of computational methods. Since this work involves specialized quantum chemistry concepts, we provide a brief glossary of methods and key concepts in Appendix \ref{complemento_info} to assist readers unfamiliar with these techniques.

Electron correlation is typically classified into two main types: static and dynamic, each playing a crucial role in electronic structure calculations. Dynamic correlation describes the instantaneous interactions between electrons, particularly in cases where electrons occupy different spatial orbitals. This is frequently referred to as a near-degeneracy effect, significant in systems where various orbitals share similar energy levels. For example, in helium, electron correlation is predominantly dynamic, while in the $H_{2}$ molecule at the dissociation limit, correlation is entirely static, where the bonding and anti-bonding molecular orbitals become degenerate. At equilibrium in $H_{2}$, correlation primarily reflects dynamic behavior, similar to helium, yet it shifts to a static form as bond length increases. Likewise, the beryllium atom exhibits both static and dynamic correlations.

Despite the challenge of clearly separating these two correlation types, they provide a useful conceptual framework for examining correlation effects. Notably, advancements in static correlation treatments include calculating energies using the full-configuration interaction method for Hamiltonians encompassing up to 100 orbitals \cite{ref_23_intro}. In contrast, addressing dynamic correlation remains a significant challenge. Traditional DMRG methods, for instance, often fall short, requiring combination with perturbation theory, and configuration interaction (CI) algorithms.

This integration poses substantial computational demands due to the expanded virtual orbital space and higher orders of reduced density matrices (RDM). Additionally, the combination of perturbative theory with density functional theory (DFT) is another promising direction, though its accuracy hinges on the choice of the functional \cite{ref_35_intro, ref_39_intro, ref_41_intro}.

In this work, we explore methods that explicitly incorporate electronic distances into the wavefunction \cite{ref_44_intro, ref_45_intro, ref_47_intro}, leading to a reduced orbital space and enhanced convergence. This is achieved by addressing the singularities associated with short-range Coulomb interactions. Thus, our approach uses the problem's fundamental nature as the pathway to its solution. 

Recent literature suggests that integrating short-range density functional into active orbital spaces results in compact, stable configurations \cite{ref_41_intro}. While F12-based algorithms have been implemented in single-reference theories, their application to multi-reference theory remains relatively unexplored. The term F12 refers to explicitly correlated methods in quantum chemistry, where the correlation function \(F(r_{12})\) explicitly depends on the interelectronic distance \(r_{12}\), improving the description of electron correlation and accelerating basis set convergence. To investigate further, we explore the transcorrelation approach initially proposed by Boys and Handy \cite{ref70}. 

The transcorrelation method (TC) models the wavefunction as the product of a CI wavefunction and a Jastrow factor - represented by the $J$ parameter, which incorporates electronic correlations \cite{ref_54_intro}. This method uses a similarity transformation on the Hamiltonian, yielding a more complex form known as the transcorrelated Hamiltonian. This transformed Hamiltonian can then be addressed with standard numerical methods for electronic Hamiltonians. However, the transcorrelation method, despite yielding a more compact wavefunction for its right eigenvector, is not widely adopted because of two primary challenges:
\begin{itemize}
    \item[1] Non-Hermitian nature of the transcorrelated Hamiltonian, which complicates classical and quantum variational algorithms as it disrupts the variational principle, particularly in establishing a lower bound.
    
    \item[2] Introduction of three-body interactions, which require additional Gaussian integral computations. These interactions also necessitate advanced measurement schemes to capture three-body fermionic reduced density matrices (RDMs), adding complexity to the transcorrelated Hamiltonian.
\end{itemize}

DMRG is widely recognized as a standard variational algorithm in various research areas, particularly in quantum chemistry, for addressing strongly correlated one-dimensional systems. Its instrumental role lies in tackling static correlation and complex electronic structures in extensive active spaces.  A major challenge in applying DMRG to ab initio systems is managing static and dynamic correlations. Numerous enhancements to DMRG have been introduced, often integrating additional methods or refinements, including active space solvers, Coupled Cluster (CC) techniques \cite{Cizek1966, Coester1958}, Configuration Interaction (CI) \cite{Sherrill1999}, perturbation theory \cite{Moller1934}, DFT \cite{Hohenberg1964, Kohn1965}, and TC \cite{ref_55_section2}. 

The considerable advancements in DMRG methodology are primarily attributed to the development of the matrix product state (MPS) framework and its associated matrix product operator (MPO). This foundational improvement has endowed DMRG with a rigorous mathematical basis, significantly enhancing the algorithm's capabilities. Moreover, it has catalyzed the exploration of a broader array of tensor network states,  notably including the development of tree tensor network states and projected entangled pair states \cite{ref_30_mpo, ref_32_mpo}. 

In its contemporary form, grounded in the MPS and MPO framework and combined with the variational principle, the DMRG algorithm serves as an invaluable tool for obtaining ground-state energies and evaluating many-body correlations.  In the classical variational algorithm presented here, as in other DMRG applications, a central objective involves constructing an MPO representation for the targeted operator. This MPO representation provides essential input for subsequent DMRG stages, specifically within many-body fermionic Hamiltonians. Here, operators fall primarily into two categories. 

The first category comprises analytical operators, such as the ab initio electronic Hamiltonian. Operators in this category are usually decomposed into a Sum of Products (SOP) formulation, enabling a systematic approach within the DMRG framework. 

The second category encompasses operators with higher complexity that lack a straightforward analytical representation.  An example is the potential operator for real molecules, characterized by N-potential energy surfaces (PES). For smaller molecules with multiple atoms, these potential operators can be accurately derived through a detailed process of globally constructing, fitting, and interpolating ab initio data points  \cite{ref_36_mpo}, allowing for precise modeling of projected entangled pair states (PEPS) for these molecular systems.

Our study focuses on operators that can be represented by SOP MPO. Constructing the most compact MPO feasible for a given operator is a key element of this approach, as compact MPOs significantly reduce computational demands.

In quantum chemistry, the prevalent manual approach to MPO construction involves symbolically designing each MPO by hand. This process entails examining the recurrence relations between neighboring sites \cite{ref_49_mpo}. A technique known as the complementary operator method \cite{ref_50_mpo} is often employed to achieve a more compact MPO, particularly for operators with long-range interactions, such as the ab initio electronic Hamiltonian. Optimizing the MPO structure with these methods is crucial for managing the operators’ computational complexity. 

Despite the benefits of manual MPO design, it lacks automation and requires custom redesign for each operator. An alternative method employs MPO compression, achieved through techniques such as Singular Value Decomposition or elimination of linearly dependent terms \cite{ref_51_mpo}. Although this method is widely used in standard libraries like ITensor \cite{itensor}, it has limitations, such as the inability to predict the numerical error introduced during compression. Furthermore, compressing large systems requires substantial computational time. 

Another strategy, rarely used in fermionic systems, is based on finite state automata, which can effectively mimic the operator’s interaction terms \cite{ref_28_mpo}. While finite-state automata are relatively straightforward to construct for regular lattices with short-range interactions, long-range interactions increase complexity, limiting their applicability to transcorrelated Hamiltonians. 

In our work, we adopt a methodology for constructing MPO inputs for transcorrelated Fermi-Hubbard model that builds upon the approach by Jiajun Ren et al. \cite{bipartite}. This approach enables the incorporation of generic Hamiltonians beyond ab initio cases and automatically generates exact, uncompressed MPOs that respect the system's symmetries and are efficient in exploring Hamiltonians which have three-body fermionic terms, as shown in this work. This algorithm brings several advantages to both classical and quantum variational algorithms based on tensor network methods:  i) Generality: it applies to all types of operators that have an analytical SOP form. ii) Automation: conversion from symbolic operator strings to MPOs is fully automated.  iii) Optimality:  the generated MPO achieves "optimal" regarding compactness  iv) Symbolic nature: the symbolic process eliminates numerical errors. 

To implement this, the method employs a bipartite graph theory framework, providing an efficient, robust approach to MPO construction in DMRG applications. This represents a significant advancement, streamlining the construction process and enhancing computational accuracy.

\subsection{Conceptual Clarifications and Terminology}

Here we briefly clarify certain terminologies used extensively in this paper to avoid potential confusion for readers unfamiliar with concepts from theoretical physics or quantum chemistry.

\subsubsection{Compactification in Theoretical Physics}

In theoretical physics, the term \textbf{compactification} typically refers to a procedure used to reduce the apparent number of dimensions of a theory, particularly common in high-energy physics, string theory, and related fields. Compactification generally involves the following steps:

\begin{itemize}
\item Starting from a higher-dimensional theory.
\item Suppose that the extra dimensions are \textbf{compactified}, or arranged in small, typically unobservable spaces.
\item Obtaining an effective theory that appears lower-dimensional.
\end{itemize}

Thus, the compactified dimensions remain hidden from the experiments at currently accessible energy scales. In string theory, compactification plays a fundamental role, since string theories are initially formulated in higher-dimensional spacetime (10 or 11 dimensions, depending on the theory).

To reconcile this with the observable four-dimensional universe, additional spatial dimensions are compactified. Important references on compactification in string theory include \cite{Green1987, Polchinski1998_2, Candelas1985, Witten1985, Becker2007,  Witten1995}. Common compactification manifolds in string theory are:

\begin{itemize}
    \item \textbf{Calabi-Yau manifolds} \cite{Candelas1985}: Complex shapes widely employed in compactifications from 10 dimensions down to our observed 4-dimensional spacetime.
    \item \textbf{Toroidal compactification (Tori)} \cite{Polchinski1998_2}: Simpler compactifications where extra dimensions form loops.
\end{itemize}

These compactifications are not merely mathematical conveniences. They critically influence the physical properties of the lower-dimensional theory, such as particle spectra, symmetries, and coupling constants.

In quantum chemistry or quantum many-body physics, the concept of \textbf{compactification} can also appear metaphorically in this work. In this context, compactification refers to the reduction of the complexity or effective dimensionality of electronic wavefunctions, making them more computationally tractable while retaining their essential physical characteristics.

\subsubsection{Error Mitigation in Quantum Computation and EMTC-DMRG}

The term \textbf{error mitigation} in the context of our work has a distinct meaning compared to its typical use in quantum computing. In this section, we clarify these differences explicitly.

In quantum computing, \textbf{error mitigation} generally refers to techniques aimed at reducing or suppressing errors in quantum computations without incurring the substantial overhead associated with full quantum error correction. Typical sources of errors include noise, decoherence, gate imperfections, or measurement inaccuracies ~\cite{temme2017error, endo2021hybrid}. The prominent error mitigation strategies used in quantum computing are:

\begin{itemize}
    \item Zero-noise extrapolation ~\cite{temme2017error, endo2021hybrid,Li2017} ;
    \item Readout error mitigation ~\cite{Bravyi2021};
    \item Quasi-probability methods  ~\cite{endo2021hybrid}
\end{itemize}

These techniques aim to enhance the reliability of quantum computations in the presence of imperfect physical implementations. In contrast, the term \textbf{error mitigation} as used in our paper, \textbf{EMTC-DMRG}, has a different conceptual context. Specifically, EMTC-DMRG explicitly focuses on mitigating intrinsic difficulties arising from the \textbf{non-Hermitian nature} of the transcorrelated Hamiltonian.

The transcorrelation (TC) method applies a similarity transformation to the Hamiltonian method to embed electron-electron correlations explicitly. A crucial consequence of this transformation is that the resulting Hamiltonian becomes \textbf{non-Hermitian} ~\cite{ref_55_section2, baiardi}. This non-Hermiticity introduces significant computational challenges.  

Numerical instabilities due to non-Hermitian eigenvalue problems, especially when employing iterative algorithms such as the Davidson solver. Besides, difficulties in achieving numerical stability and reliable convergence when applying algorithms such as the DMRG.

Therefore, in EMTC-DMRG, \textbf{error mitigation} specifically denotes strategies devised to manage or compensate for these intrinsic algorithmic challenges rather than physical quantum hardware noise. The key methods used to achieve this goal are the following:

\begin{itemize}
    \item Analytical and symbolic Matrix Product Operator (MPO) constructions to minimize numerical errors. Utilization of graph-theoretic algorithms (e.g., Hopcroft-Karp) for optimal MPO construction, thus reducing computational complexity \cite{bipartite}.
    \item Using TI- DMRG with usual Davidson solver to avoid Trotter errors \cite{davidson1975iterative}.
\end{itemize}

Additionally, optimal MPO parameters and the careful application of the Davidson solver ensure stability against numerical instabilities. In summary, EMTC-DMRG's approach to error mitigation:

\begin{itemize}
    \item Does not focus on hardware-induced quantum noise.
    \item Addresses numerical instabilities and approximation errors intrinsic to the non-Hermitian transformation involved.
    \item Employs a combination of existing techniques to ensure accuracy and computational stability.
\end{itemize}

Thus, while the term \textbf{error mitigation} typically relates to quantum computational hardware issues, our work adapts this concept to the context of algorithmic and mathematical instabilities specific to the transcorrelated methodology presented herein.

\subsection{Related Works}

It is common to encounter the spin-adapted DMRG algorithm and its optimization within MPS framework for Hermitian Hamiltonians. In our work, we extend the focus to general non-Hermitian Hamiltonians generated by transcorrelation, where the Hamiltonian does not alter the spectrum of the original Hamiltonian.

To the best of our knowledge, the literature includes two recent methodologies that propose variants of DMRG to solve TC Hamiltonians.

The first methodology, proposed by \cite{baiardi}, is named TC-DMRG. It involves an approach that employs an exact MPO, where the orbitals are optimized using partitioning and Fiedler ordering — a graph optimization strategy \cite{fiedler}. The algorithm consists of optimizing the MPS with a time-independent DMRG (TI-DMRG) and then applying the imaginary time evolution DMRG in subsequent steps. This is necessary because the target Hamiltonian breaks the variational principle. In this approach, the TC Fermi-Hubbard model, analytically derived by \cite{dobrautz}, was implemented.

The second methodology, proposed by \cite{TC_DMRG_Alavi_liao2023density}, focuses on using TI-DMRG and modifying the Davidson solver to create a generalized Davidson solver capable of handling the non-Hermitian ab initio Hamiltonian. This approach explored molecular systems using a numerically approximated TC ab initio Hamiltonian. Their findings demonstrate that implementing the TC-DMRG methodology using real numbers is indeed feasible. By employing a non-Hermitian iterative solver as referenced in \cite{ref80}, they deviate from exact diagonalization. Instead, orthogonal real trial vectors are constructed, and the real effective Hamiltonian matrix is projected onto this space. Consequently, the subspace matrix may yield complex eigenvalues and eigenvectors. Using a real-number-based non-Hermitian Davidson solver, the imaginary components of the solution are discarded—a step that can introduce numerical instability and hinder convergence. This technique was tested to solve the ground and excited states of some diatomic molecules efficiently.

\subsection{Main Contributions}

Thus, we introduce a novel variant of TC-DMRG, termed  \textit{Error-Mitigated Transcorrelated DMRG} (EMTC-DMRG). This method integrates multiple techniques, centered around the following foundational steps:
\begin{algorithm}[H]
\begin{doublespace}
\caption{EMTC - DMRG Procedure}
\begin{algorithmic}[1]
\vspace{0.9em}
\State Make a similarity transformation into its transcorrelated version \cite{dobrautz}
\vspace{0.9em}
\State Optimize the transcorrelated fermionic Hamiltonian using the Hopcroft-Karp algorithm \cite{bipartite}

\vspace{0.9em}
\State Decompose the TC Hamiltonian into an uncompressed and analytical MPO form \cite{bipartite}
\vspace{0.9em}
\State Encode the compact MPO from fermions to qubits using the Verstraete-Cirac map \cite{Verstraete_2005}
\vspace{0.9em}
\State Choose the proper transcorrelated parameter via the FCIQMC method \cite{dobrautz}
\vspace{0.9em}
\State Estimate the initial bond dimension
\vspace{0.9em}
\State Use time-independent DMRG (TI-DMRG) 
\vspace{0.9em}
\State Initialize the MPS using a right-projected or random wavefunction 
\vspace{0.9em}
\State Run the code and check the ground state energy value
\vspace{0.9em}
\end{algorithmic}
\end{doublespace}
\end{algorithm}

Through these modifications, the EMTC-DMRG algorithm has been tailored to compute ground-state energies for a two-dimensional fermionic many-body system with periodic boundary conditions. This customized approach has yielded substantial results, including reduction of long-range interactions, compactification of the fermionic wavefunction, entanglement minimization, and resource optimization by reducing the bond dimension across all cases considered.

\subsubsection{Structure of this work}

Our paper serves as a tutorial and review on the electron correlation problem and explicitly correlated methods. Additionally, it proposes a new research initiative, within which we present original results for one of the algorithms that form the foundation of this research direction. It is organized as follows:  Section \ref{1} presents a new research initiative that aims to connect several scientific domains to solve problems related to fermionic systems. Section \ref{2} provides a foundational overview of electron correlation and explicitly correlated methods, which readers familiar with these concepts may skip. Section \ref{3} outlines the key components of our methodology. Section \ref{4} presents the main results, and Section \ref{5} presents discussions that substantiate our method's efficiency. 
Our focus lies primarily on examining the EMTC-DMRG convergence behavior, comparing the number of sweeps and ground-state energy with conventional DMRG and other TC-DMRG approaches. This analysis is limited to the transcorrelated Fermi-Hubbard Hamiltonian in real representation.  Conclusions and future perspectives are discussed in Section \ref{6}.

\section{ \label{1}  New Research Initiative: It from Qubit and Bootstrap for Chemistry}

In this section, we organize our discussion into a summary of two major and contemporary scientific programs, highlighting some of the most fundamental contributions. Inevitably, many important names and works will be omitted, as our aim here is to provide a concise overview. Our perspective is also influenced by our own scientific background.
 
Throughout the history of physics, foundational progress has often been achieved by unifying distinct theoretical frameworks or connecting different domains of knowledge. In high-energy physics, Steven Weinberg, one of the leading figures of theoretical physics, was not merely concerned with solving individual problems—he was a builder of \textbf{unifying frameworks}. His pioneering work on the unification of weak and electromagnetic interactions \cite{Weinberg1967} exemplified an approach where seemingly disparate phenomena were shown to emerge from deeper, underlying principles. As Howard Georgi noted, Weinberg’s genius lay in his pursuit of \textit{"the general picture rather than specific models"} \cite{Georgi2021}, seeking overarching principles that could tie together different domains of physics.

It is worth mentioning  here one  influential example of bridging different research areas occurred in condensed matter physics. In this field, Xiao-Gang Wen revolutionized our understanding by introducing the concept of \textbf{topological order}, a classification of quantum phases beyond Landau’s symmetry-breaking paradigm. Wen’s pioneering work demonstrated that quantum phases could be characterized not by local order parameters but by global topological properties, leading to new understanding of ground state degeneracy and fractional charge in quantum Hall systems \cite{wen1989}.

He expanded this concept to general quantum many-body systems, introducing the notion of \textbf{quantum orders}, which are characterized by patterns of long-range quantum entanglement and emergent gauge symmetries \cite{wen1990}. In particular, to explain the fractional quantum Hall effect, Wen applied Chern-Simons theory—a topological quantum field theory (TQFT)—revealing that fractional charge and statistics in quantum Hall states arise from the topological properties of Chern-Simons gauge fields \cite{wen1992}. Collaborating with A. Zee, Wen further classified hierarchical fractional quantum Hall states using this formalism \cite{wenzee1992}.
Wen extended his ideas to quantum spin liquids, proposing that they host emergent gauge fields and fractionalized excitations, akin to phenomena observed in quantum chromodynamics (QCD). This led to the development of quantum orders as a new paradigm for classifying quantum phases beyond symmetry breaking \cite{wen2002}. These ideas were elaborated in his book, which provided a comprehensive review of quantum orders and emergent gauge fields \cite{wen2004}.

Inspired by string theory and quantum gravity, Wen proposed a unifying framework called \textbf{string-net condensation}, where collective excitations of fluctuating strings in a lattice give rise to emergent gauge fields and particles. This model explains the emergence of photons and fermions as low-energy excitations, providing a condensed matter interpretation of gauge bosons and fermions \cite{levinwen2005}. He later unified these descriptions using tensor category theory, bridging the gap between algebraic structures and physical observables.

Wen’s ideas have also significantly impacted \textbf{topological quantum computation}, particularly through the use of non-Abelian anyons to implement fault-tolerant quantum logic gates. This approach uses the braiding operations of anyons, which form a representation of the braid group, enabling robust quantum gates protected by topology \cite{wen2010}. Recently, Wen has expanded his framework using higher category theory and topological quantum field theory to provide a unified description of all topological phases, generalizing Chern-Simons theories and deepening the mathematical foundation of quantum orders \cite{wen2023}.

Xiao-Gang Wen’s work not only revolutionized condensed matter physics but also bridged it with high energy physics through field theory, gauge theory, and topological quantum field theory. His innovative ideas continue to inspire research in quantum information, topological materials, and quantum gravity. Crucially, Wen’s approach demonstrates the power of interdisciplinary thinking—by integrating tools from high-energy physics, he created new paradigms for understanding complex condensed matter systems.

Inspired by this philosophy of interdisciplinary synthesis, we propose a program called the \textbf{It from Qubit and Bootstrap for Chemistry}, that seeks to construct a \textbf{new interdisciplinary bridge}, connecting quantum information theory, conformal bootstrap, and quantum chemistry. Just as the typical theoretical physics tradition demonstrated as legendary figure such as Weinberg (sought to bring together the weak and electromagnetic forces into a single framework) and  Wen unified topological field theories with condensed matter physics. Inspired by several examples in science, we aim to merge insights from high-energy physics and quantum computation to develop a \textbf{systematic methodology} for understanding strongly correlated systems and \textbf{ab initio} Hamiltonians in chemistry.

This work is motivated by the recognition that while \textbf{holographic principles, tensor networks, and bootstrap techniques} have been extensively explored in high-energy physics and even in condensed matter, their applications in chemistry and how we devise quantum algorithms are largely untapped. By building upon these ideas, we seek to establish a \textbf{new paradigm} for quantum chemistry algorithms that is rigorous, general, and applicable across multiple computational architectures—ranging from classical methods to near-term and post-NISQ quantum algorithms. Our approach is based on three foundational pillars:
\begin{itemize}
    \item \textbf{It from Qubit:} Building on the idea that fundamental structures in physics emerge from quantum information, we explore how entanglement-based methods and quantum algorithms can provide new insights into the behavior of chemical systems \cite{Preskill2018}.
    \item \textbf{Conformal Bootstrap:} Inspired by the bootstrap philosophy, which extracts deep constraints from symmetry and consistency conditions, we seek to apply similar techniques to constrain the possible wavefunctions and electronic structures in chemistry \cite{Poland2019}.
    \item \textbf{Transcorrelation and Many-Body Compactification:} Recognizing that strongly correlated quantum systems often require more efficient representations, we integrate \textbf{transcorrelation techniques} to compact  wavefunction descriptions and improve computational scalability \cite{ref_55_section2, Kong2012}.
\end{itemize}

By designing a program that systematically integrates these elements, we follow in the spirit of Weinberg—not merely proposing isolated models but rather \textbf{constructing a robust theoretical} foundation that can drive new advances in chemistry, quantum computing, materials science, and condensed matter physics. This approach continues Wen's idea that all these domains should mutually inform and enrich each other.

\subsubsection*{It from qubit}

Before the emergence of the \textbf{it from qubit} paradigm, foundational contributions from physicists like Stephen Hawking and Roger Penrose shaped our understanding of the relationship between quantum mechanics, spacetime, and gravity. Hawking’s discovery of black hole radiation and his work on black hole entropy highlighted a deep connection between quantum theory and gravitational systems, suggesting that any unifying framework would need to account for information and thermodynamic properties \cite{hawking1975,hawking1976}. 

Penrose’s contributions to the theory of singularities and the structure of spacetime, as well as his philosophical inquiries into the quantum foundations of the universe, provided a mathematical and conceptual backdrop against which modern holographic and quantum information-based approaches developed \cite{penrose1965,penrose1969}. While neither Hawking nor Penrose directly engaged with the holographic principle or quantum error correction, their groundbreaking work established many of the puzzles—such as the nature of information loss in black holes—that the \textbf{it from qubit} program seeks to resolve.

The phrase \emph{“it from qubit”} emerged from a growing intersection of theoretical physics and quantum information science. Broadly speaking, it encapsulates the idea that the foundations of spacetime and gravity might be best understood through the lens of quantum entanglement, quantum error correction, and quantum information theory. This research direction is part of a broader effort to unify quantum mechanics with general relativity, a challenge that has persisted for nearly a century.

One of the earliest milestones in this research direction was the seminal paper by \cite{maldacena1997} on the Anti-de Sitter/Conformal Field Theory (AdS/CFT) correspondence. Often referred to as the “holographic principle,” this work proposed that a quantum field theory on the boundary of a space (CFT) could describe the gravitational dynamics in the bulk (AdS). While not explicitly about “it from qubit”, Maldacena’s insight laid the groundwork for understanding spacetime and gravity as emergent phenomena encoded in quantum degrees of freedom.

The notion that quantum entanglement plays a fundamental role in spacetime emerged more clearly in the 2000s. In particular, the work of \cite{vanraamsdonk2010} and collaborators suggested that the geometric connectedness of spacetime could be related to the pattern of quantum entanglement in the boundary theory. This culminated in a 2010 paper by Van Raamsdonk, which argued that increasing the entanglement between regions of the boundary theory leads to the formation of a connected spacetime in the bulk.

In parallel, developments in quantum information science—especially in quantum error correction and tensor network methods—began to influence our understanding of holography. By the mid-2010s, researchers realized that the structure of quantum entanglement in AdS/CFT could be mapped onto error-correcting codes. These codes protect quantum information in a way that mirrors how bulk gravitational information is encoded redundantly in the boundary theory.

Key papers by \cite{harlow2013}, \cite{hayden2007}, and others drew explicit connections between AdS/CFT and quantum error correction, showing how bulk operators (corresponding to spacetime regions) can be reconstructed from the boundary state. This was further formalized by researchers like John Preskill, Brian Swingle, and Patrick Hayden, who introduced ideas from tensor networks—graphical representations of quantum states that naturally encode entanglement structure—to describe aspects of the AdS/CFT correspondence \cite{swingle2012}.

The phrase \emph{“it from qubit”} was popularized around this time as an umbrella term capturing the intuition that quantum information—qubits and their entanglement—is the “it” (spacetime, gravity) in a holographic universe.

The “it from qubit” line of inquiry is still very much active. Current research explores the emergence of time, the role of complexity in holographic dualities, and the application of more advanced quantum information techniques—such as quantum error-correcting codes and tensor networks—to better understand the nature of spacetime and black holes \cite{swingle2012}.

In summary, “it from qubit” is a research program that seeks to derive the fabric of spacetime, gravity, and perhaps even fundamental physics itself from the principles of quantum information. It builds on groundbreaking work in holography, quantum error correction, and entanglement, and has been championed by leading figures in theoretical physics.

\subsubsection*{Conformal Bootstrap}

The conformal bootstrap program represents a powerful framework for studying quantum field theories (QFTs) that are invariant under the conformal group. First proposed in the 1970s, the bootstrap philosophy is built around the idea that consistency conditions—symmetry constraints and fundamental physical principles—can completely determine the dynamics of a conformal field theory (CFT) without relying on a traditional Lagrangian description \cite{polyakov1974,bpz1984}.

Conformal field theories appear naturally at critical points of statistical systems, where scale invariance enhances to full conformal invariance. In high-energy physics, they describe the fixed points of the renormalization group, including theories like $\mathcal{N}=4$ supersymmetric Yang-Mills in four dimensions or the two-dimensional minimal models classified by Belavin, Polyakov, and Zamolodchikov \cite{bpz1984}.

At its heart, the conformal bootstrap is about self-consistency. Instead of starting with a specific set of field equations or interactions, it begins with general principles:
\begin{enumerate}
    \item \textbf{Conformal Symmetry:} The structure of the conformal group severely constrains the possible correlation functions in a CFT. The scaling dimensions of fields, their spin, and their operator product expansion (OPE) coefficients must satisfy specific algebraic relations.
    \item \textbf{Crossing Symmetry:} Four-point correlation functions in a CFT can be decomposed into conformal blocks that encode contributions from operators in the spectrum. Crossing symmetry requires that different ways of decomposing these correlators yield consistent results, leading to a set of highly constraining nonlinear equations.
    \item \textbf{Unitarity and Positivity:} The CFT spectrum must respect unitarity bounds, ensuring that certain scaling dimensions are positive and that correlation functions behave properly under Hermitian conjugation \cite{simmons-duffin2019}.
\end{enumerate}

The philosophy of the bootstrap is that these constraints, when combined with computational methods, can fully determine a CFT’s spectrum and OPE coefficients without any direct reference to an action or Lagrangian.

\subsubsection*{1. Early Beginnings (1970s–1980s):}
The foundational ideas were laid out by Alexander Polyakov in 1974, who proposed the use of crossing symmetry and the conformal group to constrain correlation functions \cite{polyakov1974}. The BPZ paper in 1984 (Belavin, Polyakov, and Zamolodchikov) demonstrated the power of these ideas in two dimensions, where the Virasoro algebra further simplifies the bootstrap equations \cite{bpz1984}. They solved an infinite class of two-dimensional CFTs (the minimal models), showing how symmetry alone can fix both the spectrum and OPE coefficients.

\subsubsection*{2. Difficulties in Higher Dimensions (1980s–2000s):}
Outside of two dimensions, the conformal bootstrap initially faced challenges due to the increased complexity of conformal blocks and the lack of extended symmetry algebras. During this time, progress was limited, and most work focused on specific models or perturbative approaches.

\subsubsection*{3. Revival and Numerical Bootstrap (2008–present):}
The modern era of the conformal bootstrap began with the realization that numerical methods could be used to systematically explore the constraints of crossing symmetry and unitarity. Pioneering work by Rychkov, Rattazzi, Vichi, and others demonstrated that even in higher dimensions, these constraints can carve out an “allowed” region in parameter space, often leading to sharp predictions for critical exponents \cite{rattazzi2008}.

The development of highly efficient algorithms for computing conformal blocks, as well as the introduction of semidefinite programming techniques, led to a dramatic increase in the power of the bootstrap. For example, the numerical bootstrap has been applied to the three-dimensional Ising model, producing precise predictions for its critical exponents that rival or surpass the best Monte Carlo methods \cite{kos2016}.

\subsubsection*{4. Extensions and Applications (2010s–present):}
The conformal bootstrap is now applied far beyond the original context:
\begin{itemize}
    \item \textbf{Supersymmetric Theories:} The bootstrap is used to study supersymmetric CFTs, including $\mathcal{N}=4$ super-Yang-Mills and $\mathcal{N}=1$ theories in various dimensions. The added constraints from supersymmetry often simplify the equations and lead to new insights.
    \item \textbf{Higher-Point Functions:} While early bootstrap studies focused on four-point functions, modern techniques now tackle higher-point functions, providing more detailed data on the spectrum and OPE coefficients.
    \item \textbf{Analytic Bootstrap:} A complementary approach to numerical methods, the analytic bootstrap uses techniques like the lightcone expansion and large spin perturbation theory to extract information about CFT data in certain limits \cite{simmons-duffin2019}.
    \item \textbf{AdS/CFT Connections:} In the context of holography, the bootstrap provides a way to determine CFT data that correspond to bulk gravitational theories. This has implications for understanding the string theory landscape and quantum gravity.
\end{itemize}

The conformal bootstrap embodies a paradigm shift in theoretical physics. It demonstrates that symmetry principles and consistency conditions, when applied rigorously, can replace traditional model-building. This is particularly powerful because it sidesteps the need for a Lagrangian description, instead relying on fundamental properties of spacetime symmetry, unitarity, and causality.

The bootstrap program is also philosophically appealing because it unifies approaches from mathematics and physics. By systematically classifying possible CFTs, it bridges the gap between algebraic structures (such as conformal algebras) and physical observables. It provides a clear, almost Platonic framework: the laws of physics emerge not from arbitrary choices of equations, but from deep, immutable symmetries and logical consistency.

The conformal bootstrap is both a framework and a philosophy. By relying purely on symmetry and consistency conditions, it has become a powerful tool for exploring the nonperturbative landscape of quantum field theories. From its early roots in two dimensions \cite{bpz1984} to its modern applications in higher-dimensional CFTs, supersymmetry, and AdS/CFT, the bootstrap has continually expanded our understanding of what symmetry alone can determine.

\subsection{Program Structure}
The program is organized into four interconnected building blocks, each addressing a specific aspect of the quantum chemistry problem which comprises classical and quantum variational and non-variational algorithms, such as:
\begin{itemize}
    \item I. Variational Classical Algorithm: Error-Mitigated Transcorrelated DMRG;
    \item II. Hybrid Variational Algorithm: Transcorrelated Compressed Quantum Circuit;
    \item III. Non-Variational Classical and Quantum Algorithm: Transcorrelated Bootstrap Algorithm;
    \item IV. Non-Variational Classical and Quantum Algorithm: Transcorrelated Conformal Bootstrap Algorithm
 \end{itemize}   

\subsection{Transcorrelation and Its Renaissance in Quantum Chemistry}

The \textbf{It from Qubit and Bootstrap for Chemistry} program builds upon a renaissance of interest in transcorrelation (TC) techniques, recognizing their potential to address challenges in quantum chemistry. Originating as a method to transform the many-body Hamiltonian into a form requiring only up to three-body integrals, the TC method faced early difficulties due to its non-Hermitian nature. While initial research \cite{complemento_boys1, complemento_boys2, Handy1971} highlighted these challenges, subsequent studies \cite{complemento_10,complemento_11_Tenno_2000, complemento_12_Tenno2000b, complemento_13_Hino2002, complemento_14, complemento_15, complemento_16, complemento_17, complemento_18} demonstrated the technique's value when integrated with quantum chemical methods like Møller–Plesset perturbation theory and the linearized coupled-cluster approach.

The TC method’s application to strongly correlated systems—especially within the context of \textit{ab initio} Hamiltonians—gained renewed attention in recent years. Advances in explicitly correlated R12/F12 methods \cite{complemento_14,complemento_19_KLOPPER198717, Complemento_20, Complemento_21_KLOPPER1991583, Complemento_22_NOGA1992497, Complemento_23_SHIOZAKI2009131} and computational frameworks such as full configuration interaction quantum Monte Carlo (FCIQMC) \cite{complemento_28, ref1_intro, complemento_31} have underscored the importance of TC techniques. Despite the challenges posed by the method’s non-Hermitian nature, its ability to improve basis-set convergence and compact the representation of wavefunctions has inspired a second wave of innovation, including its adaptation to coupled-cluster and hybrid variational algorithms \cite{complemento_34, complemento_35}.

The program positions itself at the forefront of this renaissance by integrating TC techniques into a broader interdisciplinary initiative. Specifically, the TC framework serves as a foundation for developing variational and non-variational algorithms tailored to the challenges of strongly correlated systems. By leveraging insights from quantum information theory, tensor networks, and conformal bootstrap principles, the program aims to redefine the role of TC methods in quantum chemistry.

\subsection{Transcorrelation as a Key Element of Program Philosophy}

Central to the program’s vision is the use of TC methods to enhance algorithmic efficiency, scalability, and accuracy across classical, hybrid, and quantum approaches. The ability of TC methods to elevate the effective level of underlying correlation techniques aligns seamlessly with the program’s goal of compacting the fermionic many-body wavefunction. By addressing challenges such as non-variational behavior and error cancellation through innovations in Jastrow factor selection and parameter optimization, the program offers a pathway to extend TC methodologies into new domains of quantum chemistry and physics.

This interdisciplinary approach ensures that the \textbf{It from Qubit and Bootstrap for Chemistry} program contributes not only to the renaissance of TC methods but also to their evolution. By combining concepts from diverse areas of physics, the program seeks to establish TC techniques as a cornerstone of modern quantum chemistry, providing robust solutions for \textit{ab initio} systems and beyond.

Our entire toolbox, titled \textbf{Towards Compacting the Fermionic Many-Body Wavefunction}, is part of a broader initiative to efficiently manage fermionic many-body wavefunctions. 

\begin{figure}[htbp]
    \centering
    \includegraphics[width=0.7\textwidth]{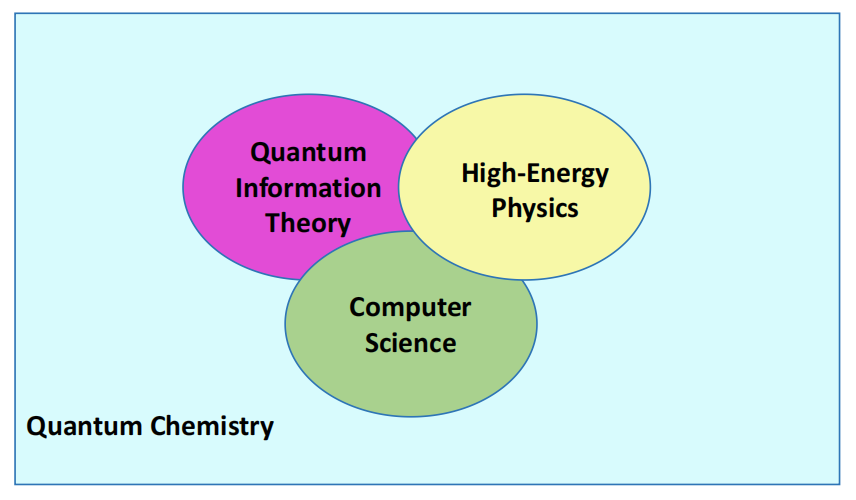} 
    \label{fig:venn}
\end{figure}

This initiative is designed to provide the foundation for advanced algorithms tailored to classical, NISQ, and post-NISQ era computing. The primary objective is to address ab initio Hamiltonians using new methodologies or hybrid approaches. For the first paper of our program, we propose a classical variational algorithm as the first building block of our toolbox. 

The program \textbf{It from Qubit and Bootstrap for Chemistry} seeks to redefine quantum chemistry by merging ideas from quantum information theory, tensor networks, and conformal field theory. Its four-block structure ensures a systematic approach to developing innovative algorithms, offering new ways to solve long standing challenges in quantum chemistry.

\section{\label{2}Theoretical Framework}

\subsection{The Electron Correlated Problem}

To begin our exploration of electron correlation, we must first consider the influence of the chosen basis set, particularly within molecular systems. Electron correlation arises as the difference between the standard probability density, \(\rho(\mathbf{r})\), and the conditional probability density, \(\rho(\mathbf{r}|\mathbf{r}')\), of finding an electron at a specific point \(\mathbf{r}\). The probability density is defined as the diagonal of the density operator, averaged over \(n-1\) electrons, summing over all spin degrees of freedom:
\begin{equation}
\rho(\mathbf{r}) = \frac{1}{n} \sum_{i=1}^{n} \rho_{i}(\mathbf{r}),
\end{equation}
\begin{equation}
\rho_{i}(\mathbf{r}) = \int \left|\psi(\mathbf{r}_{1}, \mathbf{r}_{2}, \cdots, \mathbf{r}_{n})\right|^{2} dr_{1} \cdots dr_{i-1} \, dr_{i+1} \cdots dr_{n}.
\end{equation}
For a two-electron system, the pair probability density is expressed as:
\begin{equation}
\rho(\mathbf{r}, \mathbf{r}') = \frac{1}{n(n-1)} \sum_{i \neq j}^{n} \rho_{ij}(\mathbf{r}, \mathbf{r}'), 
\end{equation}
\begin{equation}
\rho_{ij}(\mathbf{r}_{i}, \mathbf{r}_{j}) = \int \left|\psi(\mathbf{r}_{1}, \mathbf{r}_{2}, \cdots, \mathbf{r}_{n})\right|^{2} dr_{1} \cdots dr_{i-1} \, dr_{i+1} \cdots dr_{j-1} \, dr_{j+1} \cdots dr_{n}.   
\end{equation}
In the absence of correlation, the probability densities satisfy the relation
\begin{equation}
\rho(\mathbf{r}, \mathbf{r}') = \rho(\mathbf{r}) \rho(\mathbf{r}').
\end{equation}
This allows us to express the conditional probability density as: 
\begin{equation}
\rho(\mathbf{r}|\mathbf{r}') = \frac{\rho(\mathbf{r}, \mathbf{r}')}{\rho(\mathbf{r}')} .
\end{equation}
Thus, in the uncorrelated case, 
\begin{equation}
\rho(\mathbf{r}|\mathbf{r}') = \rho(\mathbf{r}),
\end{equation}
and with correlation, we have
\begin{equation}
\rho(\mathbf{r}|\mathbf{r}') \neq \rho(\mathbf{r}).  
\end{equation}
The origin of electron correlation can be attributed to two primary effects:
\begin{itemize}
    \item[1] Fermi Correlation: Due to electron indistinguishability and their adherence to Fermi-Dirac statistics, the wavefunction must change sign under the permutation of any two electrons.
    \item[2] Coulomb Correlation: The Coulomb interaction generally reduces the probability of two electrons being in close proximity.
\end{itemize}
Fermi correlation is incorporated through Slater determinants, which form fully antisymmetrized products of orbitals. A standard starting point for electronic structure calculations is the Hartree-Fock method \cite{hartree1928wave, fock1930naheansatz}, where the wavefunction is represented as a single determinant. For the ground state of a two-electron atom, the HF wavefunction is given by:
\begin{equation}
\psi_{HF}(\mathbf{r}_{1}, \mathbf{r}_{2}) = \phi(\mathbf{r}_{1}) \phi(\mathbf{r}_{2}) \frac{1}{\sqrt{2}} (\alpha_{1}\beta_{2} - \alpha_{2}\beta_{1}).    
\end{equation}
This wavefunction satisfies the conditions:
\begin{align}
\rho(\mathbf{r}) = |\phi(\mathbf{r})|^{2},
\end{align}
\begin{align}
\rho(\mathbf{r}, \mathbf{r}') = |\phi(\mathbf{r})|^{2} |\phi(\mathbf{r}')|^{2},   
\end{align}
\begin{align}
\rho(\mathbf{r}|\mathbf{r}') = |\phi(\mathbf{r})|^{2}. 
\end{align}
For the lowest-energy triplet state, however, we find: 
\begin{align}
\psi_{HF}(\mathbf{r}_{1}, \mathbf{r}_{2}) = \frac{1}{\sqrt{2}} [\phi_{1}(\mathbf{r}_{1}) \phi_{2}(\mathbf{r}_{2}) - \phi_{1}(\mathbf{r}_{2}) \phi_{2}(\mathbf{r}_{1})] \alpha_{1} \alpha_{2}.  
\end{align}
Further analysis reveals that the conditions above do not hold in this case, indicating that the wavefunction is correlated. According to Löwdin \cite{lowdin}, electron correlation is quantified as the difference between the exact non-relativistic Born-Oppenheimer energy and the HF energy:
\begin{equation}
E_{corr} \equiv E_{exact} - E_{HF} \leq 0.
\end{equation}
Since the HF wavefunction accounts for Fermi correlation, the term \textbf{electron correlation} in quantum chemistry is typically used to refer specifically to Coulomb correlation. Electron correlation can be described by various approaches, including DFT, QMC, and wavefunction theory with known analytical forms \cite{Kong2012}. In this context, we focus on wavefunction theory.

To understand how electron correlation affects the electronic wavefunction in atoms, we examine the ground state of the helium atom. This is an example of correlation between two electrons orbiting a common nucleus within the same spatial scale. The HF wavefunction for this state is statistically uncorrelated. The exact wavefunction can be expressed as:
\begin{equation}
\psi_{exact}(\mathbf{r}_{1}, \mathbf{r}_{2}) = \psi_{HF}(\mathbf{r}_{1}, \mathbf{r}_{2}) + \psi_{corr}(\mathbf{r}_{1}, \mathbf{r}_{2}),   
\end{equation}
where \(\psi_{corr}\) is orthogonal to \(\psi_{HF}\) and encodes the correlation effects not captured at the HF level. The exact wavefunction in this form is characterized by intermediate normalization: 
\begin{equation}
\left\langle \psi_{HF} \mid \psi_{exact} \right\rangle = 1,    
\end{equation}
a feature that is convenient for many-body expansions, as it facilitates the inclusion of correlation effects in a systematic manner. To gain insight into changes in the wavefunction due to electron correlation, we define:
\begin{equation}
\psi_{corr}(\mathbf{r}_{1}|\mathbf{r}_{2}) \equiv \frac{\psi_{corr}(\mathbf{r}_{1}, \mathbf{r}_{2})}{\phi(\mathbf{r}_{1})}.    
\end{equation}
This function, \(\psi_{corr}(\mathbf{r}_{1}|\mathbf{r}_{2})\), provides a measure of the correlation effects that are missing in the Hartree-Fock wavefunction, capturing the response of one electron to the presence of another. By examining this conditional form, we can more directly assess the spatial influence of correlations on the electronic structure.

\subsubsection{Configuration Interaction Wavefunction for Electron Pairs}

The CI expansion of an \(n\)-electron wavefunction represents a linear combination of Slater determinants constructed from a complete orthogonal set of orbitals \(\{\phi_{i}(\mathbf{r})\}\). The exact spin-free wavefunction for the ground state of helium, for instance, can be expressed as:
\begin{equation}
\psi_{exact}(\mathbf{r}_{1}, \mathbf{r}_{2}) = \sum_{ij} c_{ij} \phi_{i}(\mathbf{r}_{1}) \phi_{j}(\mathbf{r}_{2}),    
\end{equation}
where \(\phi_{i}\) are spin-free orbitals and For every \(i, j\), we have \(c_{ij} = c_{ji}\). Assuming this set includes the HF orbital \(\phi(\mathbf{r})\), it is also possible to represent the correlation wavefunction in CI form, as related to the spectral theorem, though this approach becomes less straightforward for systems with more than two electrons.  Importantly, the construction of the CI expansion is independent of any basis set incompleteness.

\subsubsection{Cusp Conditions}

The mathematician Tosio Kato \cite{complemento_kato}, in his study of the properties of many-body Schrödinger equations, demonstrated that wavefunctions with discrete spectra are continuous and possess bounded first derivatives, except at Coulomb singularities. At these singularities, such as the coalescence of two particles, the wavefunction's derivative exhibits a discontinuity, known as the \textbf{cusp} condition. For example, when two electrons \(i\) and \(j\) approach one another, the wavefunction's discontinuity at their coalescence point can be expressed as:

\begin{equation}
\lim_{r_{ij} \to 0} \left(\frac{\partial \psi(\cdots, \mathbf{r}_{i}, \cdots, \mathbf{r}_{j}, \cdots)}{\partial r_{ij}}\right) = \frac{1}{2} \psi\left(\cdots, \frac{\mathbf{r}_{i} + \mathbf{r}_{j}}{2}, \cdots, \frac{\mathbf{r}_{i} + \mathbf{r}_{j}}{2}, \cdots\right).
\label{cusp}
\end{equation}

Here:
\begin{itemize}
    \item \(r_{ij} = |\mathbf{r}_i - \mathbf{r}_j|\) is the inter-particle distance between electrons \(i\) and \(j\).
    \item \(\frac{\partial}{\partial r_{ij}}\) denotes differentiation with respect to this inter-particle distance, which depends on the relative positions of particles \(i\) and \(j\).
\end{itemize}

The expression evaluates the wavefunction's radial derivative as the electrons \(i\) and \(j\) approach one another (i.e., \(r_{ij} \to 0\)). The wavefunction on the right-hand side is evaluated at the midpoint of the two particle positions, \(\mathbf{r}_i\) and \(\mathbf{r}_j\), through spherical averaging over a hypersphere where \(r_{ij}\) is constant and \(\mathbf{r}_i + \mathbf{r}_j\) is constant.

Similar cusp conditions arise at other Coulomb singularities, such as electron-nucleus coalescence. These conditions can also be derived from the three-dimensional Schrödinger equation by solving the radial and angular parts using spherical harmonics and considering the Coulomb interaction's behavior near singularities.

\subsubsection{Explicitly Correlated Wavefunctions}

The idea of using interelectronic distances to construct efficient wavefunctions dates back to 1927, when Slater  proposed a wavefunction for two-electron atoms that matched the Rydberg limit, where both electrons are close to the nucleus \cite{slater}. In this limit, Slater proposed a wavefunction behaving as \(\sim e^{-2(r_1 + r_2)}\), assuming no penetration corrections. Near the atomic core, his analysis suggested a modified form \(\sim e^{-2(r_1 + r_2) + r_{12}/2}\), in line with the Rydberg limit.  The resulting wavefunction,
\begin{equation}
\psi_{\text{exact}}(\mathbf{r}_1, \mathbf{r}_2, \mathbf{r}_{12}) = e^{-2(r_1 + r_2) + r_{12}/2},    
\end{equation}
provided an improved approximation, yielding an energy close to the observed value of approximately \(2.856 E_h\) - $E_{h}$ is the HF reference energy \cite{James1933}. Slater’s early analyses, published in subsequent works, also examined this wavefunction's utility in predicting properties such as helium's magnetic susceptibility \cite{ref35_section2}. 
Simultaneously, Hylleraas developed his groundbreaking variational wavefunction for the helium ground state, achieving remarkable accuracy within 0.1 eV (see \cite{Hylleraas1928}) of the observed value. His wavefunction, formulated in terms of \(\mathbf{r_1}\), \(\mathbf{r_2}\), and \(\cos\theta_{12}\), is described in polar coordinates as \(\mathbf{r_{12}} = \{r_{12}, \theta_{12}, \phi_{12}\}\). It included both even and odd powers of \(\mathbf{r_1}\) and \(\mathbf{r_2}\), while only even powers of \(r_{12}\), as seen from
\begin{equation}
r_{12}^2 = (\mathbf{r}_1 - \mathbf{r}_2) \cdot (\mathbf{r}_1 - \mathbf{r}_2) = r_1^2 + r_2^2 - 2r_1r_2\cos\theta_{12}.    \end{equation}
This relationship led to a more compact wavefunction with a linear dependence on \(\mathbf{r_{12}}\),
\begin{equation}
\psi_1(\mathbf{r_1}, \mathbf{r_2},\mathbf{r_{12}}) = N(1 + c_1(r_1 - r_2)^2 + c_2r_{12})e^{-\alpha(r_1 + r_2)},    
\end{equation}
where parameter  $c_{1}$, $c_{2}$, and $\alpha$ were determined variationally ($c_{1}=0.130815$, $c_{2}=0.291786$, $\alpha=-1.81607$ and the normalization constant is $1.330839$ \cite{Kong2012}. Further error reduction was achieved by including higher powers of \(\mathbf{r}_1 \pm \mathbf{r}_2\) and \(\mathbf{r_{12}}\). The rapid asymptotic convergence of Hylleraas’s wavefunction, compared to the slower convergence of CI-type expansions, can be attributed to the inclusion of linear and odd powers of \(\mathbf{r_{12}}\). CI-type wavefunctions experience slow convergence near the cusp, primarily due to the absence of odd \(\mathbf{r_{12}}\) powers and the global support of Slater determinants. Wavefunctions with local support could more effectively capture cusp features even without these odd-power terms. While \(\mathbf{r_{12}}\) terms are essential for compact wavefunctions, reoptimizing orbitals is also key to achieving this compactification. 

Explicitly correlated wavefunctions are vital for accurately modeling the cusps in electronic wavefunctions, enabling rapid decay of basis set errors in many-electron systems. The high dimensionality of integrals involving interelectronic distances (\(\mathbf{r_{12}}\)) has traditionally hindered practical applications of concepts introduced by Slater and Hylleraas. Overcoming these barriers required innovative techniques for efficiently evaluating such integrals, a topic we explore in subsequent sections. 

\subsubsection{Chronological order}

Extending Hylleraas’s ideas to more complex systems is conceptually straightforward but technically challenging due to the complexity of high-dimensional integrals. Consequently, explicitly correlated wavefunctions have seen primary application in high-precision atomic and molecular physics. Notable advancements include the development of R12 methods, which have extended the utility of explicitly correlated methods to general molecular systems  \cite{ref10_section2}. The evolution of these methods, in chronological order, includes: 
\begin{enumerate}[label=\alph*.]
    \item Wavefunctions for two-electron systems.
    \item Wavefunctions for \(n\)-electron systems.
\end{enumerate}
Within the realm of \(n\)-electron systems, several notable methodologies have been developed:
\begin{enumerate}
    \item \textbf{Hylleraas CI method} - Enhances the variational approach by systematically incorporating interelectronic distances \cite{hyl_60}. 
    \item \textbf{Explicitly correlated Gaussian methods} - Utilizes Gaussian functions to directly model electron correlations \cite{Boys1960_gaussian,Singer1960_gaussian}.
    \item \textbf{Many-body Gaussian geminal methods} - Extends Gaussian methods by applying geminal pairing for electron groups \cite{Kong2012}.
    \item \textbf{Transcorrelated method} - Applies a non-unitary transformation to the Hamiltonian, simplifying the treatment of electron correlation \cite{ref_55_section2}.
\end{enumerate}
This work focuses primarily on the transcorrelated method, exploring its distinctive contributions to the field and providing our contribution.

\subsubsection{Explicitly Correlated Wavefunctions for \(n\)-Electron Systems}

Applying explicitly correlated techniques to molecular systems with more than two electrons presents a key challenge: the need to evaluate numerous computationally intensive many-electron integrals. Even when restricting each \(n\)-electron basis function to depend on only one of the interelectronic distances, up to four integrals are required. Addressing these integrals is central to advancing the application of explicitly correlated wavefunctions. Over the years, several methods have been developed to tackle this issue:
\begin{enumerate}

    \item Weak orthogonality functionals can be used to avoid some high-dimensional integrals \cite{ref57_section2}.
     
    \item For atoms or specific explicitly correlated basis functions, it is possible to evaluate integrals exactly, as in the case of explicitly correlated Gaussian functions, which allow analytical evaluation for multi-electron molecules \cite{ref53_section2}.

    \item Stochastic evaluation of \(n\)-electron integrals is utilized in variational QMC methods. Diffusion and other true QMC methods, along with Nakatsuji’s local Schrödinger equation (LSE) method, are closely related approaches  \cite{ref58_section2,ref_59_section2}.
    
    \item Resolution of identity (RI) is used in R12 methods \cite{ref_39_section2} to simplify three- and four-electron integrals, reducing them to two-electron integrals. R12 methods incorporate many of the previously mentioned techniques \cite{Kong2012}. 

    \item The similarity transformation of the Hamiltonian analytically removes Coulomb singularities (transcorrelated method). This yields a more complex Hamiltonian with three-electron terms, though the resulting cusp-less wavefunctions can be effectively described using Slater determinants \cite{ref70, ref57_section2}.
\end{enumerate}

\subsection{Transcorrelated Method}

\subsubsection{The Transcorrelated Method and its Extensions}

Solving the Schr\"odinger equation for many-electron systems poses significant challenges, even for relatively simple molecules such as  \(\mathrm{H}_2\). To make these calculations tractable, the continuous real-space Hamiltonian is often projected onto a finite Hilbert space, defined by a basis set typically composed of \(M\) atomic or molecular spin-orbitals.  This vector space includes all possible anti-symmetrized products (Slater determinants) of \(N\) electrons in \(M\) spin-orbitals, ensuring the Pauli exclusion principle is satisfied. The anti-symmetry requirement is naturally handled through the formalism of second quantization, where the wavefunction is expressed as: 
\begin{equation}
\left|\psi\right\rangle = \sum_\kappa \alpha_\kappa \left|\kappa\right\rangle,
\end{equation}
where $\alpha_\kappa$ are complex coefficients, and $\left|\kappa\right\rangle$ represents Slater determinants expressed conveniently as Fock occupation number vectors: 
\begin{equation}
\left|\kappa\right\rangle = \left|i_{M-1}, \cdots, i_j, \cdots, i_0\right\rangle,
\end{equation}
where $i_j = 1$ indicates that spin-orbital $j$ is occupied, and $i_j = 0$ indicates it is empty.
Standard quantum chemistry methods—such as HF, CC, CI, and multiconfigurational self-consistent field theory (MCSCF)—approximate the electronic Schrödinger equation by using wavefunctions of this form. These methods either restrict the number of determinants in the expansion or include all terms but apply approximations to the coefficients $\alpha_\kappa$.

However, as early as the 1920s, it was observed that expanding wavefunctions as linear combinations of Slater determinants results in slow convergence of the true eigenvalues and eigenfunctions. This convergence issue arises from the failure of basis expansions to accurately resolve the sharp features of the wavefunction at electron-electron coalescence points. These “cusp conditions”  were formalized by Kato \cite{ref_18_section2_tc}, and wavefunctions constructed purely from single-particle orbitals do not satisfy these conditions \cite{ref_19_section2_tc}. The TC method, originally developed by Boys and collaborators and inspired by earlier work by Hirschfelder \cite{ref_8_section2_tc, complemento_boys1, ref_23_section2_tc}, addresses these issues by transforming the Hamiltonian rather than the wavefunction to capture electron correlation. This idea draws a parallel to the Heisenberg picture of quantum mechanics, where operators evolve in time, as opposed to the Schrödinger picture, where wavefunctions change and operators remain static. The transformation is defined by applying a correlation factor to the wavefunction:
\begin{equation}
\left|\psi\right\rangle = e^{\sum_{i<j} f(\mathbf{r}_i, \mathbf{r}_j)} \left|\phi\right\rangle,
\end{equation}
where $f(\mathbf{r}_i, \mathbf{r}_j)$ is a symmetric real function representing the correlation between electrons $i$ and $j$, and $\left|\phi\right\rangle$ is a reference wavefunction without explicit inter-electronic distance dependence. The real-space Schrödinger equation becomes:
\begin{equation}
H \left|\psi_\kappa\right\rangle = E_\kappa \left|\psi_\kappa\right\rangle,
\end{equation}
 being $E_{\kappa}$ is the eigenenergy which after applying the transcorrelation transformation, becomes: 
\begin{equation}
H e^{\widehat{g}} \left|\phi_\kappa\right\rangle = E_\kappa e^{\widehat{g}} \left|\phi_\kappa\right\rangle,
\end{equation}
or equivalently,
\begin{equation}
H' \left|\phi_\kappa\right\rangle = E_\kappa \left|\phi_\kappa\right\rangle,
\end{equation}
where $H' = e^{-\widehat{g}} H e^{\widehat{g}}$ is the transcorrelated Hamiltonian. Thus, while the explicitly correlated wavefunction $\left|\psi_\kappa\right\rangle$ is an eigenstate of the original Hamiltonian, the eigenvalue can also be obtained by solving for the simpler wavefunction $\left|\phi_\kappa\right\rangle$, which is an eigenstate of $H'$.  The following form for the correlation function $f_{ij}$:
\begin{equation}
f_{ij}=\sum_{k}c_{k}g_{k}(\mathbf{r}_{i},\mathbf{r}_{j})+\sum_{\mu}d_{\mu}\phi_{\mu}(\mathbf{r}_{i}),
\end{equation}
where $g_{k}$ is known as Gaussian explicitly correlated basis function (a Gaussian geminal for two electrons) and $\phi_{\mu}$ is an atomic orbital (non-orthogonal orbital) \cite{gaussian_53, Boys1960_gaussian}. Being that this primitive Gaussian geminal reads
\begin{equation}
g_{k}(\mathbf{r}_{i},\mathbf{r}_{j})=\exp\left(-\alpha_{1k}|\mathbf{r}_{i}-A_{k}|^{2}-\alpha_{2k}|\mathbf{r}_{j}-B_{k}|^{2}-\gamma_{k}r^{2}\right),
\end{equation}
where the $\alpha_{ik}$, $\gamma_{k}$, $B_{k}$, and $A_{k}$ are non-linear parameters which need to be optimized variationally \cite{Kong2012}. This equation is a product of two s-type Gaussian functions with a Gaussian correlation factor $\exp\left(-\gamma_{k}r^{2}\right)$. Note that the Gaussian geminals do not satisfy the cusp condition \eqref{cusp} but its linear combination can be approximated of the shape of the electronic wavefunction near the cusp due the dependence of the explicitly interelectronic distance.
Handy successfully applied this method to small molecular systems, such as $H_{2}$, achieving accurate results with a small number of parameters \cite{ref_78_section2_tc}.

\subsubsection{Limitations of the Transcorrelated Hamiltonian}

While the transcorrelated method reduces computational complexity, it also introduces challenges due to the nonunitary nature of the transformation \( e^{\widehat{g}} \), resulting in a non-Hermitian TC Hamiltonian. This non-Hermitian property presents several issues: 
\begin{enumerate}
    \item The absence of a strict variational lower bound on the ground-state energy.
    \item Different right- and left-hand eigenvectors, complicating the calculation of observables beyond energy.
    \item Non-orthogonal right-hand (and left-hand) eigenvectors, increasing the complexity of their construction.
\end{enumerate}
The non-Hermitian nature of the TC Hamiltonian complicates the evaluation of observables beyond energy, as these calculations rely on both left- and right-hand eigenvectors. The expectation value of an observable \( \hat{\mathcal{O}} \) is given by:
\begin{equation}
\langle O \rangle = \langle \psi | \hat{\mathcal{O}} | \psi \rangle = \langle \phi_\kappa | e^{\widehat{g}} \hat{\mathcal{O}} e^{\widehat{g}} | \phi_\kappa \rangle.
\end{equation}
However, since \( e^{\widehat{g}} \hat{\mathcal{O}} e^{\widehat{g}} \) generally does not have a terminating Baker-Campbell-Hausdorff (BCH) expansion, an approximation is often applied: 
\begin{equation}
\hat{\mathcal{O'}} = e^{-\widehat{g}} \hat{\mathcal{O}} e^{\widehat{g}},
\end{equation}
which does terminate. This leads to the following expression for the observable: 
\begin{equation}
\langle O \rangle = \langle \widetilde{\phi}_\kappa | \hat{\mathcal{O'}} | \phi_\kappa \rangle,
\end{equation}
where \( \langle \widetilde{\phi}_\kappa | = \langle \phi_\kappa | e^{2\widehat{g}} \) represents the left-hand eigenvector of the TC Hamiltonian. The inclusion of additional dynamic correlation in the left-hand eigenvector makes its construction from a single-particle basis expansion more complex \cite{ref_19_section2_tc}.

\subsubsection{Practical Considerations }

When expanded in real space through the BCH series, the transcorrelated Hamiltonian introduces two- and three-body terms that significantly alter its structure: 
\begin{equation}
H' = H + [H, \hat{g}] + \frac{1}{2} [[H, \hat{g}], \hat{g}],
\end{equation}
which can be expressed as: 
\begin{equation}
H' = H - \sum_i \left(\frac{1}{2} \nabla_i^2 \hat{g} + (\nabla_i \hat{g}) \nabla_i + \frac{1}{2} (\nabla_i \hat{g})^2 \right).
\end{equation}
The TC method has been extended to various models, such as the Fermi-Hubbard model, where Dobrautz et al. \cite{dobrautz} and Tsuneyuki \cite{ref_24_section2_tc} applied a Gutzwiller factor: 
\begin{equation}
H' = \left(e^{-J \sum_i n_{i,\uparrow} n_{i,\downarrow}}\right) H \left(e^{-J \sum_j n_{j,\uparrow} n_{j,\downarrow}}\right),
\end{equation}
where $J$ is the Jastrow factor, \( n_{i,\sigma} \) is the number operator for the spin-lattice site indexed by \( i \) and \( \sigma \). This transformation suppresses double occupancies on lattice sites, producing a ``compactification``  of the right-hand eigenvectors, thus simplifying their approximation. However, as right-hand eigenvectors become easier to approximate, the left-hand eigenvectors incorporate additional dynamic correlation, complicating their construction from a single-particle basis expansion. 

Early implementations of the TC method involved optimizing parameters in the Jastrow function, resulting in wavefunctions based on single-particle orbitals optimized through self-consistent equations (TC-SCF)  \cite{ref_8_section2_tc, complemento_boys1}. However, the non-Hermitian property of the TC Hamiltonian complicates the application of the Rayleigh-Ritz variational principle, posing difficulties in confirming convergence  \cite{Handy1971}.

Handy proposed minimizing the variance of the energy to approximate the ground state, but this approach introduces additional challenges \cite{Handy1971}. Minimizing the variance may yield less accurate energy estimates than direct energy minimization. Furthermore, projecting the TC Hamiltonian onto a single-particle basis results in up to \( \mathcal{O}(M^6) \) terms, compared to \( \mathcal{O}(M^4) \) terms in the unmodified Hamiltonian, due to the inclusion of three-electron terms.

\subsubsection{Recent Advances in the Transcorrelated Method}

In recent years, substantial progress has been made in addressing the computational challenges associated with the TC method. Ten-no \cite{ref_26_section2_tc} and Hino \cite{ref_27_section2_tc} fixed the form of the Jastrow function and compensated for errors by expanding the reference wavefunction as a sum of Slater determinants. Another approach combined TC-SCF with variational Monte Carlo, using energy variance as the optimization criterion  \cite{ref_24_section2_tc}. Luo \cite{ref_32_section2_tc, ref_33_section2_tc} improved upon these methods by discarding terms linear in \( g \) to obtain a Hermitian operator.

Luo and Alavi \cite{ref_11_section2_tc} further advanced the TC method by integrating it with the full configuration interaction quantum Monte Carlo (FCIQMC) approach. This method employs a wavefunction with a frozen Jastrow term and a Slater determinant expansion optimized using FCIQMC. Since FCIQMC is related to the imaginary-time evolution of the state, the ground state can be determined by evolving the system for a sufficiently long time:
\begin{equation}
\left|\psi_0\right\rangle = \lim_{\tau \to \infty} e^{-H \tau} \left|\psi\right\rangle,
\end{equation}
and similarly,
\begin{equation}
\left|\phi_0\right\rangle = \lim_{\tau \to \infty} e^{-H' \tau} \left|\phi\right\rangle.
\end{equation}
By transferring dynamic correlation from the wavefunction to the TC Hamiltonian, Luo and Alavi substantially reduced the number of walkers required in FCIQMC simulations, thereby enhancing computational efficiency.  Their approach has been successfully applied across a range of systems  \cite{ref_12_section2_tc}.
The transcorrelated method provides a versatile framework for managing the complexity of many-body quantum problems by transforming the Hamiltonian to more effectively incorporate electron correlation effects.  Despite the challenges posed by its non-Hermitian nature, including complexities in computing observables beyond energy, recent developments have substantially improved its applicability. The TC method holds considerable potential, especially in the context of quantum computational algorithms, and continues to drive progress in electronic structure theory.

\subsection{2D Lattices Mapped as 1D Chain}

The mapping of fermions onto qubits is a crucial step in quantum simulations, especially for quantum many-body systems. Fermions, due to their antisymmetric exchange statistics, exhibit non-local properties in state space. This behavior is encoded in the \textbf{Pauli exclusion principle}, which fundamentally distinguishes fermionic systems from their bosonic counterparts. Specifically, the wavefunction of a fermionic system acquires a phase of $\pi$ when two fermions are exchanged, resulting in their anticommutation properties.

To perform quantum simulations, it is necessary to represent fermionic degrees of freedom on qubits. However, fermions inherently exhibit \textbf{non-locality} in their state space due to their exchange statistics. As fermions are exchanged, the wavefunction acquires a phase factor of $-1$ when traced past an odd number of other fermions, and no phase change when traced past an even number. This phase factor reflects the \textbf{antisymmetric} property of fermions.

Despite this, \textbf{parity superselection} ensures that only states with definite fermion parity (either even or odd) can exist, preventing the measurement of relative phases due to fermion exchange. Thus, any representation of fermions on distinguishable systems, such as qubits, must introduce non-local operators, as is evident in the \textbf{Jordan-Wigner (JW) transformation} \cite{Jordan1928}.

The JW transformation provides a method for mapping fermionic creation and annihilation operators, which satisfy the canonical anticommutation relations:
\begin{equation}
\{ a_i^\dagger, a_j \} = \delta_{ij}, \qquad \{ a_i^\dagger, a_j^\dagger \} = 0, \qquad \{ a_i, a_j \} = 0,
\end{equation}
into Pauli operators on qubits. Specifically, the fermionic creation operator $a_i^\dagger$ is mapped to the following non-local operator:
\begin{equation}
a_i^\dagger \rightarrow \frac{1}{2} Z_1 Z_2 \cdots Z_{i-1} (X_i - iY_i),
\end{equation}
where $Z_j$, $X_j$, and $Y_j$ are the Pauli operators acting on qubit $j$. This transformation converts local fermionic operators, such as the creation and annihilation operators, into non-local qubit operators involving strings of Pauli-Z operators.

Under the Jordan-Wigner transformation, even local observables that conserve fermion parity, such as lattice hopping terms, are mapped to long strings of Pauli operators. 

\subsubsection{Fermion-to-Qubit Mappings: Challenges and Solutions}

Fermion-to-qubit mappings, such as the Jordan-Wigner transformation, allow a fermionic Hamiltonian $H_f$ to be mapped to a qubit Hamiltonian $H_q$. This is essential for simulating fermionic systems. After the mapping, the qubit Hamiltonian takes the form:
\begin{equation}
H_q = \sum_i H_i,
\end{equation}
where each $H_i$ is a tensor product of Pauli operators acting on different qubits. However, the \textbf{Pauli weight}—the number of qubits on which an individual term $H_i$ acts—affects the complexity of the quantum circuit required to simulate the system. High Pauli weight increases the circuit depth and complexity, directly impacting the performance of variational algorithms such as the hybrid variational algorithms \cite{LQOT_bruna}, time-dependent classical variational algorithm, and even adiabatic quantum algorithms \cite{AQA_axel}.

To reduce the computational cost, it is important to design fermion-to-qubit mappings that minimize the Pauli weight of common fermionic operators. Specifically, mappings should ensure that \textbf{geometrically local} fermionic interactions, such as those between nearby fermionic modes on a lattice, are mapped to qubit interactions that also remain local or have low Pauli weight.

\subsubsection{Fermionic Encodings: Balancing Non-Locality and Efficiency}

Quantum simulations of fermionic systems depend on efficient encodings from fermions to qubits. The \textbf{non-local structure} of fermionic Fock space requires encodings that map local fermionic operators into non-local qubit operators or use entangled subspaces to represent the fermionic degrees of freedom. Two key properties of fermionic encodings are:
\begin{itemize}
    \item \textbf{Low-weight representation} of local fermionic operators to reduce circuit depth.
    \item \textbf{Error resilience} to mitigate physical qubit errors during the quantum computation.
\end{itemize}
Despite the apparent trade-off between these two properties, recent research \cite{ JW_3_localfermionic} suggests that low-weight encodings can still exhibit error-mitigating properties, especially when the undetectable errors correspond to natural fermionic noise. For example, in certain encodings, undetectable single-qubit errors map to local fermionic phase noise, which is a type of error naturally occurring in fermionic systems. Therefore, even low-weight encodings can suppress error rates in fermionic simulations without sacrificing the efficiency of the encoding.

\subsubsection{Stabilizer Formalism for 2D Lattices and the Jordan-Wigner Transformation}

The \textbf{stabilizer formalism} is a powerful tool for encoding quantum information in subspaces of a larger Hilbert space. It is widely used in the context of quantum error correction, where a subset of qubits is restricted to the simultaneous $+1$ eigenspace of a group of commuting Pauli operators, known as the \textbf{stabilizer group}. In the case of fermionic systems, stabilizer codes offer a way to encode fermionic states into qubits while maintaining logical equivalence with the fermionic system.

\subsubsection{Stabilizer Codes and the Code Space}

Consider a quantum system composed of $N$ qubits. A stabilizer code defines a subspace, known as the \textbf{code space}, by specifying a set of stabilizer generators $\{S_1, S_2, \dots, S_k \}$, where each $S_i$ is a Pauli operator (or a tensor product of Pauli operators) acting on the $N$ qubits. These stabilizers satisfy the following properties:
\begin{itemize}
    \item[i] They commute with each other, i.e., $[S_i, S_j] = 0$ for all $i, j$.
    \item[ii] Each stabilizer has eigenvalues $\pm 1$, and the code space is defined as the subspace of the Hilbert space where all stabilizers have eigenvalue $+1$, i.e., the set of states $|\psi\rangle$ such that $S_i |\psi\rangle = |\psi\rangle$ for all $i$.
\end{itemize}
The dimension of the code space is $2^{N-k}$, where $k$ is the number of independent stabilizer generators. For fermionic systems encoded into qubits, the stabilizers impose constraints on the system, ensuring that the encoded states retain the desired fermionic properties, such as parity conservation.
For further details about this topic we refer the reader to Refs. \cite{JW_1_Derby_2021, JW_2_Chiew_2023, JW_3_localfermionic}.

\section{\label{3} Our Algorithm: EMTC-DMRG}

\subsection{Model: Transcorrelated Fermi-Hubbard Hamiltonian}
In the case study system explored here to test our algorithm, we use the transcorrelation formulation devised by \cite{dobrautz}. To compute the ground-state energy of the two-dimensional, single-band Hubbard model, we consider the Hamiltonian in real-space basis:
\begin{equation}
\hat{H} = -t \sum_{\langle ij \rangle, \sigma} a_{i,\sigma}^\dagger a_{j,\sigma} + U \sum_l n_{l,\uparrow} n_{l,\downarrow},
\end{equation}
where \(a_{i,\sigma}^{(\dagger)}\) are fermionic operators, \(n_{l,\sigma}\) the number operator, \(t\) the nearest-neighbor hopping parameter, and \(U \geq 0\) the on-site Coulomb interaction. We adopt a Gutzwiller-type Ansatz for the ground-state wavefunction:
\begin{equation}
\ket{\Psi} = g^{\hat{D}} \ket{\Phi} = e^{\hat{\tau}} \ket{\Phi}, \quad \text{where } \hat{\tau} = (ln g) \times \hat{D}  = J \sum_l n_{l,\uparrow} n_{l,\downarrow},
\end{equation}
with \(0 \leq g \leq 1\) and \(\hat{D}\) representing the double-occupancy operator. Here, \(J\) is optimized using Variational Monte Carlo (VMC), minimizing:
\begin{equation}
E_{VMC} = \min_J \frac{\braopket{\Phi_0}{e^{\hat{\tau}} \hat{H} e^{\hat{\tau}}}{\Phi_0}}{\braopket{\Phi_0}{e^{2\hat{\tau}}}{\Phi_0}}.
\end{equation}
For this study, \(\ket{\Phi}\) is expanded as a full configuration interaction (CI) wavefunction:
\begin{equation}
\ket{\Phi} = \sum_i c_i \ket{D_i},
\end{equation}
allowing us to solve the similarity transformed eigenvalue equation:
\begin{equation}
e^{-\hat{\tau}} \hat{H} e^{\hat{\tau}} \ket{\Phi} = \bar{H} \ket{\Phi} = E \ket{\Phi}.
\end{equation}
Here, \(\bar{H}\) is the similarity transformed Hamiltonian:
\begin{equation}
\bar{H} = -t \sum_{\langle ij \rangle, \sigma} e^{-\hat{\tau}} a_{i,\sigma}^\dagger a_{j,\sigma} e^{\hat{\tau}} + U \sum_l n_{l,\uparrow} n_{l,\downarrow}.
\end{equation}
Expanding the commutators using the Baker-Campbell-Hausdorff series, we find:
\begin{equation}
\hat{F}(x) = e^{-x\hat{\tau}} a_{i,\sigma}^\dagger a_{j,\sigma} e^{x\hat{\tau}},
\end{equation}
leading to:
\begin{equation}
\hat{F}(1) = a_{i,\sigma}^\dagger a_{j,\sigma} e^{J(n_{j,\bar{\sigma}} - n_{i,\bar{\sigma}})}.
\end{equation}
Here, \(\bar{\sigma}\) denotes the spin opposite to \(\sigma\), meaning \(\bar{\sigma} = \uparrow\) when \(\sigma = \downarrow\) and vice versa.
The transformed Hamiltonian becomes:
\begin{widetext}
  \begin{equation}
  \begin{split}
    \bar{H} = &-t \sum_{\langle i,j\rangle ,\sigma} a_{i,\sigma}^{\dagger} a_{j,\sigma} 
    + U \sum_{l} n_{l,\uparrow} n_{l,\downarrow} \\
    &- t \sum_{\langle i,j\rangle ,\sigma} a_{i,\sigma}^{\dagger} a_{j,\sigma} 
    \left\{ (e^J - 1) n_{j,\bar{\sigma}} + (e^{-J} - 1) n_{i,\bar{\sigma}} - 2(\cosh(J) - 1) n_{i,\bar{\sigma}} n_{j,\bar{\sigma}} \right\}.
  \end{split}
  \end{equation}   
\end{widetext}
The exponential factor in the transformed Hamiltonian can be computed to facilitate the analysis of the system in momentum space, as the physical characteristics of the low to intermediate \( U/t \) regime are not adequately described by the ansatz in real space. This necessitates the translation of the Hamiltonian into momentum space:
\begin{align}
\hat{a}_{\mathbf{r},\sigma}^{\dagger} = \frac{1}{\sqrt{M}} \sum_{\mathbf{k}} e^{-ikr} \hat{c}_{\mathbf{k},\sigma}^{\dagger},
\end{align}
 where $M$ is the lattice dimension. The exact non-Hermitian transcorrelated Hamiltonian \( \bar{H} \) in \( \mathbf{k} \) space, post-similarity transformation, is expressed as:
\begin{widetext}
\begin{align}
\begin{split}
\overline{H}_{\mathbf{k}}(J) = & -t \sum_{\mathbf{k},\sigma} \epsilon_{\mathbf{k}} n_{\mathbf{k},\sigma} \\
& + \frac{1}{M} \sum_{\mathbf{pqk},\sigma} \omega(J,\mathbf{p},\mathbf{k}) \hat{c}_{\mathbf{p-k},\sigma}^{\dagger} \hat{c}_{\mathbf{q+k},\overline{\sigma}}^{\dagger} \hat{c}_{\mathbf{q},\overline{\sigma}} \hat{c}_{\mathbf{p},\sigma} \\
& + 2t \frac{\cosh(J) - 1}{M} \sum_{\mathbf{pqskk'},\sigma} \epsilon_{\mathbf{p-k+k'}} \hat{c}_{\mathbf{p-k},\sigma}^{\dagger} \hat{c}_{\mathbf{q+k'},\overline{\sigma}}^{\dagger} \hat{c}_{\mathbf{s+k-k'},\overline{\sigma}}^{\dagger} \hat{c}_{\mathbf{s},\overline{\sigma}} \hat{c}_{\mathbf{q},\overline{\sigma}} \hat{c}_{\mathbf{p},\sigma}.
\end{split}
\end{align}
\label{kspace}
\end{widetext}
Here, \( \omega(J,\mathbf{p},\mathbf{k}) \) is defined as:
\begin{align}
\omega(J,\mathbf{p},\mathbf{k}) = \frac{U}{2} - t [ (e^J - 1) \epsilon_{\mathbf{p-k}} + (e^{-J} - 1) \epsilon_{\mathbf{p}} ].    
\end{align}
\sloppy
This exact transformation simplifies the ground-state calculation using methods like FCIQMC while maintaining the Hamiltonian's spectrum. Although \(\bar{H}\) is non-Hermitian, projective methods such as stochastic FCIQMC enable direct sampling of the ground-state energy without variational optimization. The non-Hermiticity, though challenging for observables, facilitates compact wavefunction sampling, which is beneficial for numerical studies in the intermediate correlation regime.

In this work, we do not present results using a k-space TCFH Hamiltonian. However, for studies involving larger lattices, it is necessary to modify the Hamiltonian to operate in momentum space while preserving the rest of the algorithm.

\subsection{MPOs}
\label{subsec:mpo}

The wavefunction ansatz in the DMRG method is referred to as the MPS or tensor train representation, and it is expressed as:
\begin{equation}
    |\Psi\rangle 
    = \sum_{\{b\},\{\alpha\}}
      B[1]^{\alpha_1}_{b_1} B[2]^{\alpha_2}_{b_1 b_2} \cdots
      B[N]^{\alpha_N}_{b_{N-1}} \,
      \bigl|\alpha_1 \alpha_2 \cdots \alpha_N\bigr\rangle,
    \label{eq:mps}
\end{equation}
where \(N\) denotes the number of degrees of freedom (DoFs) in the system for distinguishable particles or the number of orbitals in an electronic system. The local basis states \(\{|\alpha_i\rangle\}\) could represent a discrete variable representation (DVR) basis for nuclear motion or orbital occupation configurations for electronic systems. For spatial orbitals, \(\{|\alpha_i\rangle\} = \{\vert \mathrm{vacuum}\rangle, \vert\uparrow\rangle, \vert\downarrow\rangle, \vert\uparrow\downarrow\rangle\}\), while for spin orbitals, \(\{|\alpha_i\rangle\} = \{\vert \mathrm{vacuum}\rangle, \vert \mathrm{occupied}\rangle\}\).

The local tensors \(B[i]^{\alpha_i}_{b_{i-1} b_i}\) are interconnected via the indices \(b_i\), often referred to as virtual bonds, with bond dimension \(D\) (denoted as \(\lvert b_i \rvert\)). The indices \(\alpha_i\) are called physical bonds with dimension \(d\). A key advantage of the DMRG method is its controllable accuracy, determined by the bond dimension \(D\), which can be systematically increased.

Analogous to MPS, any operator \(\hat{Q}\) can be represented as a MPO \cite{schollwock2011density,chan2016matrix}:
\begin{equation}
    \hat{Q} = \sum_{\{v\}, \{\beta\}, \{\beta'\}}
    X[1]^{\beta'_1, \beta_1}_{v_1} \,
    X[2]^{\beta'_2, \beta_2}_{v_1 v_2} \cdots
    X[N]^{\beta'_N, \beta_N}_{v_{N-1}}
    \,\bigl|\beta'_1 \beta'_2 \cdots \beta'_N\bigr\rangle
    \bigl\langle \beta_N \beta_{N-1} \cdots \beta_1\bigr|,
    \label{eq:mpo}
\end{equation}
where \(\{X[i]^{\beta'_i, \beta_i}_{v_{i-1} v_i}\}\) are tensors defining the operator, connected by virtual bonds \(v_i\) with dimension \(M_X\). While constructing an MPO numerically using singular value decomposition (SVD) of the full operator matrix \(\mathbf{Q}_{\beta'_1\beta'_2\cdots \beta'_N,\, \beta_1\beta_2\cdots \beta_N}\) is possible, this method is computationally expensive for large systems due to exponential scaling in bond dimension. For a system with \(d\)-dimensional local basis and even \(N\), the bond dimension follows a sequence \(d^2, d^4, \cdots, d^{N-2}, d^N, d^{N-2}, \cdots, d^2\).

In practice, operators often have a SOP structure, enabling a more efficient, symbolic MPO construction:
\begin{align}
    \hat{Q} 
    &= \sum_{\{k\}} c_{k_1 k_2 \cdots k_N} 
       \prod_{i=1}^N \hat{o}_i^{k_i}, 
    \label{eq:mpo_operator_basis}\\
    &= \sum_{\{v\}, \{k\}}
       Y[1]^{k_1}_{v_1} 
       Y[2]^{k_2}_{v_1 v_2} \cdots
       Y[N]^{k_N}_{v_{N-1}} 
       \prod_{i=1}^N \hat{o}_i^{k_i},
    \label{eq:mpo_mps}\\
    &= \sum_{\{v\}} 
       \hat{Y}[1]_{v_1} 
       \hat{Y}[2]_{v_1 v_2} \cdots
       \hat{Y}[N]_{v_{N-1}},
    \label{eq:mpo_final}
\end{align}
where \(\{\hat{o}_i^{k_i}\}\) are elementary operators at each site, such as \(\{\hat{I}, \hat{p}^2, \hat{x}, \hat{x}^2\}\) for vibrational systems or \(\{\hat{I}, \hat{a}^\dagger, \hat{a}, \hat{a}^\dagger \hat{a}\}\) for electronic systems. Coefficients \(c_{k_1 k_2 \cdots k_N}\) are typically sparse, simplifying symbolic MPO construction.

\paragraph{Systematic MPO Construction via Recursion}

A systematic method for constructing an MPO leverages the recursive splitting of the system into left (L) and right (R) blocks at site \(i\). This recursion expresses \(\hat{Q}\) as:
\begin{equation}
    \hat{Q} 
    = \sum_{p=1}^P 
      \hat{Q}_{[1:i]}^p 
      \otimes 
      \hat{Q}_{[i+1:N]}^p,
    \label{eq:split}
\end{equation}
where \(\hat{Q}_{[1:i]}^p = \prod_{j=1}^i \hat{o}_j^{p}\) and \(\hat{Q}_{[i+1:N]}^p = \prod_{j=i+1}^N \hat{o}_j^{p}\) are left- and right-block operators, respectively. Overlapping terms between blocks can often be combined, reducing redundancy.

The \textbf{complementary operator technique} addresses cases where interaction terms share components, allowing compact MPO representations by merging redundant operators. For example, constructing MPOs for electronic Hamiltonians, which involve terms like
\(\sum_{pqrs} g_{pqrs}\,\hat{a}_p^\dagger \hat{a}_q^\dagger \hat{a}_r \hat{a}_s,\)
requires careful design of complementary operators to reduce bond dimensions from \(\mathcal{O}(N^4)\) to \(\mathcal{O}(N^2)\) \cite{keller2015efficient,chan2016matrix}.

The challenge in constructing efficient MPOs lies in designing complementary operators systematically, as the correlation between left and right blocks complicates automation. Optimal MPO designs remain an art, guided by experience and the specific structure of the Hamiltonian \cite{chan2016matrix}.

\subsection{MPO Construction Approach Using Bipartite Graph Theory}
\label{sec:core_alg}

We now present an automated strategy for constructing MPOs by interpreting the operator-selection problem at each bond as a minimum vertex-cover problem in a bipartite graph.  We then show that the locally optimal solution obtained by this approach is, in fact, also optimal from a global perspective.

\begin{figure}[!ht]
    \centering
    \includegraphics[width=0.8\textwidth]{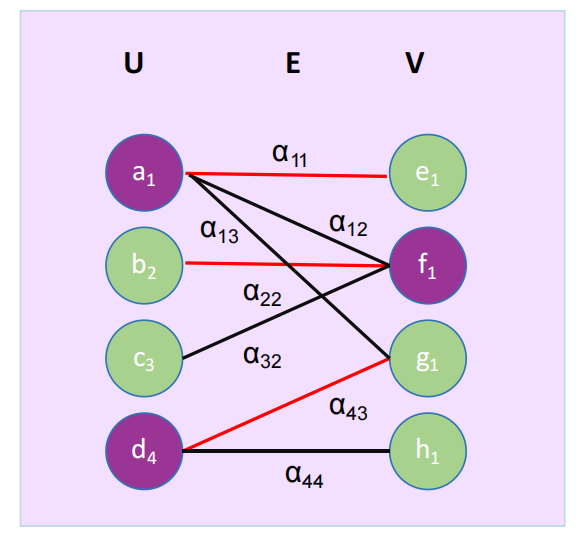} 
    \caption{The diagram illustrates the mapping of the operator
    $\hat{Q} = \alpha_{11}a_{1}e_{1} + \alpha_{12}a_{1}f_{2} + \alpha_{13}a_{1}g_{3} + \alpha_{22}b_{2}f_{2} + \alpha_{32}c_{3}f_{2} + \alpha_{42}d_{4}g_{3} + \alpha_{44}d_{4}h_{4}$ to a bipartite graph $G = (U, V, E)$ which is composed by the elements between the sets $U = {a_1, b_2, c_3, d_4}$ and $V = {e_1, f_2, g_3, h_4}$.  The vertices represents a non-redundant operator in the left and right block of the partition.  The edges connecting the vertices represent the relationships defined by weights $\alpha_{ij}$, where $i$ and $j$ denote the corresponding nodes connected by the edge. The edges shown in red form a maximum matching. The purple vertices form a maximum vertex cover.}
    \label{fig:figure2}
\end{figure}

\textbf{Mapping the Operators to a Bipartite Graph.}
After removing any duplicates among the L-block operators $\{\hat{Q}[1:i]_{r_i}\}$ and R-block operators $\{\hat{Q}[i+1:N]_{s_i}\}$ in Eq. \eqref{eq:split}, we denote the resulting unique operators by
\[
R = \{\hat{R}[1:i]_{r_i}\}, 
\quad 
S = \{\hat{S}[i+1:N]_{s_i}\}.
\]
These sets are represented by vertices in Fig. \ref{fig:figure2}.  Although each term in Eq. \eqref{eq:split} pairs exactly one L-block operator with one R-block operator, in practice the same $\hat{R}[1:i]_{r_i}$ may appear in multiple interaction terms, leading to a one-to-many mapping from $r_i$ to $s_j$.  Each of the $K$ nonzero interactions is represented as an edge in the bipartite graph $G = (R, S, E)$, with coefficient (weight) $\gamma_{r_i s_j}$.

Selecting a particular vertex in $R$ means retaining the corresponding $\hat{R}[1:i]_{r_i}$ in the L-block.  All R-block operators $\{\hat{S}[i+1:N]_{s_j}\}$ connected to that vertex by edges then combine (with their respective prefactors) into a single complementary operator for the R-block:
\[
\sum_{s_j \in \,\mathrm{edges}(r_i)} \gamma_{r_i s_j} \, \hat{S}[i+1:N]_{s_j}.
\]
Analogous logic applies if one chooses a vertex in $S$.  Consequently, to cover all interaction terms using as few retained L- or R-block operators as possible, one needs the fewest vertices covering all edges in the bipartite graph---i.e., the minimum vertex cover (blue vertices in Fig. \ref{fig:figure2}).  By K\"onig's theorem, for bipartite graphs the size of the minimum vertex cover equals the size of the maximum matching (red edges).\cite{bondy1976graph}

\paragraph{Algorithmic Steps.}
Suppose we fix a certain ordering of the sites in the DMRG chain.  To build the MPO of $\hat{Q}$ from site $1$ to $N$ (a similar procedure applies from $N$ to $1$):

\begin{enumerate}[label*=\arabic*.]
\item \label{item:step1_new}
\textbf{Obtain incoming operators at site $i$.}
  Let $\{\hat{W}[1:i-1]_{w_{i-1}}\}$ be the non-redundant set of operators (both normal and complementary) that emerge from site $i-1$.  For $i=1$, this incoming set is simply $\{\mathbb{I}\}$.  Multiply these by the local elementary operators $\{\hat{z}_i\}$ at site $i$, forming
  \[
     \{\hat{R}[1:i]_{r_i}\} 
     \;=\; 
     \{\hat{W}[1:i-1]_{w_{i-1}}\} \,\otimes\, \{\hat{z}_i\}.
  \]
  The R-block non-redundant set $\{\hat{S}[i+1:N]_{s_i}\}$ consists of all normal operators for the remaining sites.  Only those interactions with nonzero prefactors are kept.  Hence, between sites $i$ and $i+1$, we have
  \[
     \hat{Q}
     \;=\;
     \sum_{r_i,\,s_i} \gamma_{r_i,\,s_i}\,\hat{R}[1:i]_{r_i} 
          \,\otimes\, \hat{S}[i+1:N]_{s_i}.
  \]

\item \label{item:step2_new}
\textbf{Construct and solve the bipartite graph.}
  Form the bipartite graph $G = (R, S, E)$ by taking $R = \{\hat{R}[1:i]_{r_i}\}$ and $S = \{\hat{S}[i+1:N]_{s_i}\}$ as vertex sets, and introducing an edge for every nonzero $\gamma_{r_i s_i}$.  Compute a maximum matching (for example, by the Hopcroft--Karp or Hungarian algorithm \cite{hopcroft1973n,kuhn1955hungarian}), then identify the minimum vertex cover via K\"onig's theorem.  Finally, for each chosen vertex in the cover:
  \begin{enumerate}[label*=\arabic*.]
  \item If it is $r_i \in R$, we keep $\hat{R}[1:i]_{r_i}$ directly and remove its edges from the graph.
  \item If it is $s_i \in S$, we build the complementary operator
    \[
       \hat{\widetilde{R}}[1:i]_{s_i} 
       \;=\; 
       \sum_{r_i \in \,\mathrm{edges}(s_i)}
         \gamma_{r_i s_i}\,\hat{R}[1:i]_{r_i},
    \]
    retain that combination, and remove the associated edges.
  \end{enumerate}
  Removing edges ensures each interaction is counted exactly once.  When finished, the graph has no edges left.

\item \label{item:step3_new}
\textbf{Update the outgoing operators at site $i$.}
  The new set of retained operators in the L-block, 
  \[
     \{\hat{W}[1:i]_{w_i}\} 
     \;=\; 
     \bigl\{\hat{R}[1:i]_{r_i}\bigr\} 
      \,\cup\, 
     \bigl\{\hat{\widetilde{R}}[1:i]_{s_i}\bigr\},
  \]
  becomes the outgoing operator set for site $i$ and the incoming one for site $i+1$.  Using $\{\hat{W}[1:i-1]_{w_{i-1}}\}$ and $\{\hat{W}[1:i]_{w_i}\}$, we can immediately write the local symbolic MPO tensor $\hat{W}[i]$ via the relation
  \[
     \hat{W}[1:i] 
     \;=\;
     \hat{W}[1:i-1]\;\hat{W}[i].
  \]
  In practice, the prefactors $W[i]^{z_i}_{w_{i-1}\,w_i}$ appear as a transformation matrix from the operator basis 
  $\{\hat{W}[1:i-1]_{w_{i-1}}\}\otimes\{\hat{z}_i\}$
  to 
  $\{\hat{W}[1:i]_{w_i}\}$.  

\end{enumerate}

\paragraph{Local Optimality Implies Global Optimality.}
At first glance, one might worry that choosing a minimal vertex cover at each boundary only provides a local optimum.  However, we can show this procedure is also globally optimal.  Briefly, let $\gamma_{z_1 z_2 \cdots z_N}$ be reshaped into a matrix $\gamma_i$ indexed by $(z_1 \ldots z_i)$ versus $(z_{i+1} \ldots z_N)$.  This \emph{unfolding matrix} \cite{oseledets2011tensor} has rank $r_i$ and directly corresponds to the adjacency matrix for the bipartite graph at bond $i$.  By a theorem due to Lov\'asz, the maximal matching in that bipartite graph has size $r_i$ \cite{lovasz1979determinants}.  Consequently, the minimum number of operators needed at bond $i$ is $r_i$, and the above sweeping procedure does achieve this rank in each local partition.  

Numerically oriented approaches such as SVD-based compression \cite{hubig2017generic} attempt to approximate these ideal ranks but can be hampered by floating-point errors.  In contrast, the bipartite graph method here is exact and maintains sparsity in the MPO.  Moreover, it can handle symmetries by assigning quantum numbers to normal and complementary operators, and it applies equally to constructing MPS if a wavefunction in Fock-space form is already known.

Finally, note that for systems with inhomogeneous interaction patterns, the ordering of degrees of freedom still influences the ultimate MPO size.  No known polynomial-time algorithm universally provides the optimal ordering in terms of minimal bond dimensions.  Nevertheless, this question is typically of lower priority than the well-known site-ordering problem for achieving accurate DMRG convergence \cite{legeza2003optimizing,moritz2005convergence}.

\subsection{Fermionic exact MPO}

    This section reviews the matrix product operator (MPO) decomposition as originally proposed by Jiajun et al. in \cite{bipartite}. Their groundbreaking work provides the theoretical foundation for the discussions that follow. Here, we aim to extend their descriptions to enhance pedagogical clarity and make the complex concepts more accessible to readers who may not have a specialized background in this area. For a thorough understanding of the foundational methods, readers are encouraged to refer to Jiajun et al.'s original publication.

In our algorithm, we used an automatic exact MPO method. Here we will focus on the application of this methodology in the representation of the electronic Hamiltonian, focusing on its decomposition and the calculation of bond dimensions for efficient computational simulation. The electronic Hamiltonian is expressed as:
\begin{equation}
    \hat{H}_{el} = \sum_{p,q=1}^{N} h_{pq} \hat{a}_{p}^{\dagger} \hat{a}_{q} + \frac{1}{2} \sum_{p,q,r,s=1}^{N} \nu_{pqrs} \hat{a}_{p}^{\dagger} \hat{a}_{q}^{\dagger} \hat{a}_{r} \hat{a}_{s}.
\end{equation}
This Hamiltonian contains two key components as the one-electron terms, $h_{pq}$, which include kinetic energy and electron-nucleus attraction, and the two-electron terms, $\nu_{pqrs}$, representing electron-electron Coulomb repulsion. To construct an efficient MPO representation, the Hamiltonian is decomposed into three parts:
\begin{enumerate}
    \item Intra-block terms ($H_1$), which describe interactions within either the left or right blocks of the DMRG partition.
    \item Inter-block two-electron terms ($H_2$), which describe interactions between two orbitals from the left block and two from the right block.
    \item Inter-block three-electron terms ($H_3$), which involve three operators from one block and one operator from the other.
\end{enumerate}
The goal is to derive the bond dimensions $M_O$ for each of these components, as they dictate the computational complexity of the MPO.
The intra-block Hamiltonian describes interactions that occur entirely within the left or right block. It can be expressed as:
\begin{equation}
\begin{aligned}
    \hat{H}_1 &= \sum_{p_{L},q_{L}=1}^{n_L} h_{p_{L}q_{L}} \hat{a}_{p_{L}}^{\dagger} \hat{a}_{q_{L}} + \frac{1}{2} \sum_{p_{L},q_{L},r_{L},s_{L}=1}^{n_L} \nu_{p_{L}q_{L}r_{L}s_{L}} \hat{a}_{p_{L}}^{\dagger} \hat{a}_{q_{L}}^{\dagger} \hat{a}_{r_{L}} \hat{a}_{s_{L}} \\
    &+ \sum_{p_{R},q_{R}=1}^{n_R} h_{p_{R}q_{R}} \hat{a}_{p_{R}}^{\dagger} \hat{a}_{q_{R}} + \frac{1}{2} \sum_{p_{R},q_{R},r_{R},s_{R}=1}^{n_R} \nu_{p_{R}q_{R}r_{R}s_{R}} \hat{a}_{p_{R}}^{\dagger} \hat{a}_{q_{R}}^{\dagger} \hat{a}_{r_{R}} \hat{a}_{s_{R}}.
\end{aligned}
\end{equation}
Since the left and right blocks do not interact in this term, the MPO requires a minimal bond dimension. At each bond, the MPO needs to represent only two possibilities such as the identity operator $I$ when no interaction is occurring, and the intra-block Hamiltonian term. Thus, the bond dimension for $H_1$ is:
\begin{equation}
    M_{O,1} = 2.
\end{equation}
This bond dimension reflects the simple structure of $H_1$, as there are no inter-block couplings involved. The second part of the Hamiltonian, $H_2$, involves two-electron interactions that couple the left and right blocks. These terms are given by:
\begin{equation}
\begin{aligned}
    \hat{H}_2 &= \sum_{p_L < q_R, r_L < s_R} -g_{p_L q_R r_L s_R} \hat{a}_{p_L}^{\dagger} \hat{a}_{r_L} \hat{a}_{q_R}^{\dagger} \hat{a}_{s_R} \\
    &+ \sum_{p_L < q_L, r_R < s_R} g_{p_L q_L r_R s_R} \hat{a}_{p_L}^{\dagger} \hat{a}_{q_L}^{\dagger} \hat{a}_{r_R} \hat{a}_{s_R} \\
    &+ \sum_{p_R < q_R, r_L < s_L} g_{p_R q_R r_L s_L} \hat{a}_{r_L} \hat{a}_{s_L} \hat{a}_{p_R}^{\dagger} \hat{a}_{q_R}^{\dagger}.
\end{aligned}
\end{equation}
Here, $p_L, r_L$ refer to orbitals in the left block, while $q_R, s_R$ refer to orbitals in the right block. These terms describe how two operators from the left block interact with two operators from the right block. To determine the bond dimension for $H_2$, we first calculate the number of operator pairs. In the left block, there are $n_L^2$ ways to choose two creation operators, and similarly, $n_R^2$ ways to choose two operators in the right block. To reduce the bond dimension, we introduce a complementary operator technique. A complementary operator, such as:
\begin{equation}
    \hat{P}_{qs} = \sum_{p, r} -g_{p_L q_R r_L s_R} \hat{a}_{p_L}^{\dagger} \hat{a}_{r_L},
\end{equation}
allows us to aggregate multiple terms acting within a block. This reduces the number of independent terms we need to store. The bond dimension for $H_2$ is thus:
\begin{equation}
    M_{O,2} = \min(n_L^2, n_R^2) + 2 \min\left(\frac{n_L(n_L-1)}{2}, \frac{n_R(n_R-1)}{2}\right).
\end{equation}
The first term, $\min(n_L^2, n_R^2)$, accounts for the number of two-electron interactions between the blocks, while the second term represents internal pairings within each block. 
The three-electron Hamiltonian term $H_3$ is more complex than the previous terms because it involves three operators in one block interacting with one operator from the other block. The full expanded form of the Hamiltonian $H_3$ is as follows:
\begin{widetext}
\begin{equation}
\begin{aligned}
    \hat{H}_3 = &\sum_{p} \hat{a}_{p_{L}}^{\dagger} \left( \sum_{q} h_{p_{L} q_{R}} \hat{a}_{q_{R}} + \sum_{qrs} g_{p_{L} q_{R} r_{R} s_{R}} \hat{a}_{q_{R}}^{\dagger} \hat{a}_{r_{R}} \hat{a}_{s_{R}} \right)  \\
    &+ \sum_{r} \hat{a}_{r_{L}} \left( \sum_{p} \frac{-1}{2} h_{s_{R} r_{L}} \hat{a}_{s_{R}}^{\dagger} + \sum_{prs} g_{p_{R} q_{R} r_{L} s_{R}} \hat{a}_{p_{R}}^{\dagger} \hat{a}_{q_{R}}^{\dagger} \hat{a}_{s_{R}} \right)  \\
    &+ \sum_{q} \left( \sum_{p} \frac{-1}{2} h_{q_{R} p_{L}} \hat{a}_{p_{L}} + \sum_{prs} g_{p_{L} q_{R} r_{L} s_{L}} \hat{a}_{p_{L}}^{\dagger} \hat{a}_{r_{L}} \hat{a}_{s_{L}} \right) \hat{a}_{q_{R}}^{\dagger}  \\
    &+ \sum_{s} \left( \sum_{r} \frac{-1}{2} h_{r_{L} s_{R}} \hat{a}_{r_{L}}^{\dagger} + \sum_{pqr} g_{p_{L} q_{L} r_{L} s_{R}} \hat{a}_{p_{L}}^{\dagger} \hat{a}_{q_{L}}^{\dagger} \hat{a}_{r_{L}} \right) \hat{a}_{s_{R}}.
\end{aligned}
\end{equation}
\end{widetext}
In this expanded form, each of the four lines represents terms where three creation/annihilation operators act within one block (either the left or the right), and a single operator acts in the other block. This complexity makes the bond dimension larger than in the two-electron case ($H_2$) because of the many possible ways to combine these operators. 

In the first row we find a term where a creation operator from the left block interacts with a sum of operators in the right block. The second and third rows represent similar interactions but with operators acting on the right or left blocks, respectively. Finally, the fourth line represents an interaction where two creation operators and one annihilation operator from the left block interact with a single operator from the right block.

Each of these terms introduces a large number of independent operator combinations, and if not carefully reduced, the bond dimension could increase significantly.

In \(H_2\), the \textbf{complementary operator technique} was employed to group terms and reduce the number of matrix elements stored in the MPO. However, due to the increased complexity in \(H_3\), this strategy alone is not sufficient, as \(H_3\) involves many different combinations of three operators acting in one block, which significantly increases the bond dimension when using complementary operators. 

To address this, an is used for \(H_3\) and instead of only applying complementary operators, the MPO construction makes use of additional grouping techniques at the boundaries of the DMRG chain, where the asymmetry between the number of left and right operators can be exploited. An alternative strategy is employed for \(H_3\), incorporating additional grouping techniques at the DMRG chain boundaries rather than relying solely on complementary operators. This approach leverages the asymmetry between the number of left and right operators.

Specifically, terms are aggregated by focusing on interactions that are primarily between one index (either from the left or right block) while minimizing the number of independent matrix elements for the three operators.  By prioritizing cases where the number of single-index operators (from the right block) is greater than that of three-index operators (from the left block), this technique optimizes MPO storage and reduces computational complexity.

The bond dimension for \(H_3\) depends on the number of ways to combine three operators 
from one block with one operator from the other block. The number of combinations of three 
operators in one block is:
\begin{equation} 
    \binom{n_L}{3} = \frac{n_L(n_L-1)(n_L-2)}{6}. 
\end{equation}
By employing this alternative strategy, the bond dimension 
is reduced from a potential \(\mathcal{O}(N^3)\) scaling to a more computationally efficient structure. The bond dimension for \(H_3\) is then given by:
\begin{equation} 
M_{O,3} = 2\min\left(\frac{n_L^2(n_L-1)}{2}, n_R\right) + 2 \min\left(n_L, \frac{n_R^2(n_R-1)}{2}\right). 
\end{equation}
This expression quantifies the possible combinations of three operators from the left block 
interacting with one operator from the right block. The strategy optimally aggregates terms, 
minimizing the number of matrix elements stored in the MPO and ensuring computational efficiency.

For the \(H_3\) term, the standard complementary operator technique does not fully suffice due to the large number of operator combinations. Therefore, a different strategy is used to reduce the bond dimension, primarily by taking advantage of the asymmetry in the interactions between the left and right blocks. By employing this strategy, the bond dimension is reduced from the expected \(\mathcal{O}(N^3)\) scaling for three-electron terms to a more manageable form, making it suitable for practical DMRG computations.

Thus, the total bond dimension \( M_O \) for the entire MPO is the sum of the contributions from \( H_1 \), \( H_2 \), and \( H_3 \). The maximum bond dimension typically occurs when the left and right blocks are of equal size (\( n_L = n_R = N/2 \)):
\begin{equation} 
    M_{O,\text{max}} = 2 \left(\frac{N}{2}\right)^2 + 3 \left(\frac{N}{2}\right) + 2.
\end{equation}
This indicates that the bond dimension scales quadratically with the number of orbitals, while complementary operators help control this growth, making the MPO representation more efficient.

The detailed explanations provided here build upon the critical insights of Jiajun et al., with the aim to broaden the educational reach and applicability of their method. By analyzing and elaborating on their original work, this section seeks to enhance comprehension for an interdisciplinary audience.

\subsection{Mapping Fermions to Qubits via Jordan-Wigner on 2D Lattices} 

In a 1D system, the \textbf{Jordan-Wigner transformation} maps fermionic operators to qubit 
operators with relative ease, although it introduces non-local 
Pauli-Z strings. However, extending the Jordan-Wigner transformation to higher-dimensional 
systems, such as 2D lattices, presents considerable challenges.
The primary complication arises from the fact that fermionic systems must maintain the 
anti-commutation relations of fermionic operators across a 2D grid, which results in non-local mappings in the qubit encoding.

\subsubsection{Jordan-Wigner Mapping in 2D} 

In a 2D lattice, each lattice site corresponds to a qubit, and the fermionic creation and 
annihilation operators must still obey the anti-commutation relations. The straightforward 
extension of the Jordan-Wigner transformation to 2D requires selecting
a specific ordering of the lattice sites, imposing a 1D path (usually a \textbf{snake-like ordering}) across the 2D lattice. 
The fermionic creation and annihilation operators at site \( i \) are then mapped similarly 
to the 1D case:
\begin{equation} 
a_i^\dagger \rightarrow \frac{1}{2} Z_1 Z_2 \cdots Z_{i-1} (X_i - iY_i),
\end{equation}
\begin{equation} 
a_i \rightarrow \frac{1}{2} Z_1 Z_2 \cdots Z_{i-1} (X_i + iY_i),
\end{equation}
where the product of \( Z \)-operators extends over all sites preceding \( i \) according to the 
chosen 1D ordering of the 2D lattice. The Jordan-Wigner transformation renders local fermionic operators, such as nearest-neighbor hopping terms, into non-local qubit operators.
For example, consider fermionic creation and annihilation operators \( a_i^\dagger \) and \( a_j \) on neighboring sites \( i \) and \( j \). The hopping term in the fermionic Hamiltonian:
\begin{equation} 
H_{\text{hop}} = a_i^\dagger a_j + a_j^\dagger a_i, 
\end{equation}
is transformed into a qubit Hamiltonian with the form:
\begin{equation} 
H_{\text{hop}} \rightarrow \frac{1}{2} (X_i X_j + Y_i Y_j) \prod_{k=i+1}^{j-1} Z_k.
\end{equation}
On a 2D lattice, the Pauli-Z string's length increases with the separation 
between \( i \) and \( j \) in the selected 1D ordering. This introduces non-local interactions, elevating the Pauli weight of the operators and making circuit implementation on quantum computers more complex.

\subsubsection{Reducing Non-Locality: The Role of Stabilizers} 

To mitigate the non-locality introduced by the Jordan-Wigner transformation in 2D, stabilizer 
codes can be employed. These codes introduce auxiliary qubits and employ stabilizers to counteract the long Pauli-Z strings, thereby reducing the Pauli weight of the encoded operators.

In the \textbf{Verstraete-Cirac (VC) encoding} \cite{Verstraete_2005}, for example, auxiliary qubits are introduced at each lattice site, and the stabilizer generators are 
defined in such a way that they cancel out the long strings of \( Z \)-operators in the mapped 
fermionic operators. Each stabilizer is constructed as a product of Pauli operators acting on the primary and auxiliary qubits, ensuring that the non-local Pauli strings appearing in the Jordan-Wigner transformation are minimized or eliminated.
For example, consider the stabilizer corresponding to a pair of neighboring lattice sites \( i \) and \( j \) that are not consecutive in the Jordan-Wigner ordering. The stabilizer generator is constructed as a product of Pauli operators acting on the auxiliary qubits:
\begin{equation} 
S_{ij} = Z_i' Z_j',
\end{equation}
where \( Z_i' \) and \( Z_j' \) are Pauli-Z operators acting on the auxiliary qubits associated with 
sites \( i \) and \( j \). The stabilizers are designed to cancel the Pauli-Z strings in the encoded fermionic operators, leading to a more localized qubit representation.
Thus, the VC encoding modifies the hopping term to the following form:
\begin{equation} 
\tilde{a}_k^\dagger \tilde{a}_i + \tilde{a}_i^\dagger \tilde{a}_k = \frac{1}{2} \left( Y_i Y_k + X_i X_k \right) Z_i' \left( \prod_{j=i+1}^{k-1} Z_j Z_j' \right),
\end{equation}
where the auxiliary qubits shorten the length of the Pauli strings, leading to a more efficient 
quantum circuit. In 2D lattices, the design of the stabilizer codes must account for the \textbf{connectivity} of the lattice. In a square lattice with nearest-neighbor interactions, each site connects to up to four neighbors. The stabilizers must be constructed to reflect this lattice structure, ensuring that the non-local Pauli strings are effectively minimized across all interactions.

For example, in the VC encoding, the stabilizers are chosen such that every edge \( (i, k) \) of the lattice that connects two non-consecutive sites in the Jordan-Wigner ordering has an associated stabilizer that cancels the long Pauli strings. This guarantees that the encoded 
fermionic operators preserve locality relative to the original lattice geometry.

The stabilizers are typically defined as pairs of encoded Majorana operators on the auxiliary 
sites. For example, the stabilizer associated with the edge between sites \( i \) and \( k \) is 
given by:
\begin{equation} 
S_{ik} = i \tilde{\mu}_{i'} \tilde{\mu}_{k'},
\end{equation}
where \( \tilde{\mu}_{i'} \) and \( \tilde{\mu}_{k'} \) are the encoded Majorana operators on the 
auxiliary qubits at sites \( i \) and \( k \), respectively. These stabilizers effectively cancel long Pauli strings in the original fermionic operators, producing 
localized qubit operators.
Fermions are typically described in physical systems as particles with continuous positions and 
momenta. However, many quantum systems can be discretized, approximating the fermions as 
occupying discrete modes or lattice sites. Thus, fermion-to-qubit mappings play a crucial role in converting these discrete fermionic modes into qubits for quantum computation. 
This is particularly useful in material science and condensed matter physics, where models like 
Fermi-Hubbard model simulate fermions on a lattice with local interactions.

The \textbf{Fermi-Hubbard model} is a central example of such lattice models and is widely used 
for studying strongly correlated fermionic systems. Defined on a discrete lattice, this model describes spin-1/2 fermions that can hop between sites and interact when occupying the same site. The Fermi-Hubbard 
Hamiltonian is given by:
\begin{equation} 
H_{\text{FH}} = -t \sum_{\langle i, j \rangle, \sigma} (c_{i\sigma}^\dagger c_{j\sigma} + 
\text{h.c.}) + U \sum_i n_{i\uparrow} n_{i\downarrow},
\end{equation}
where \( t \) is the hopping amplitude, \( U \) is the on-site interaction energy, \( c_{i\sigma}^\dagger \) 
and \( c_{i\sigma} \) are the creation and annihilation operators for fermions with spin \( \sigma \) at 
site \( i \), and \( n_{i\sigma} = c_{i\sigma}^\dagger c_{i\sigma} \) is the number operator. Each site hosts either an occupied fermion state (odd parity) or an empty state (even parity). The interactions are 
local with respect to the lattice geometry, making it well-suited for simulation via fermion-to-qubit mappings.

Fermion-to-qubit encodings are essential for simulating fermionic systems on quantum computers, where qubits represent fermionic modes and their interactions model fermionic hopping and interactions. These mappings are also fundamental for classical simulations using tensor networks, for instance. A key challenge they must address is the non-local structure of fermions, which complicates the design of efficient quantum circuits. Mappings like Jordan-Wigner and Verstraete-Cirac enable the representation 
of these lattice models on qubit-based systems, advancing the study of complex fermionic 
behavior in quantum chemistry and condensed matter physics.

The \textbf{Jordan-Wigner transformation} provides a foundation; however,
advanced encodings like the \textbf{Verstraete-Cirac} method offer practical solutions by 
reducing Pauli weight and improving computational efficiency. These improvements are especially beneficial for near-term quantum algorithms, where reducing circuit depth and mitigating physical qubit errors are crucial.

\subsubsection{Example}
In tensor network methods like MPO, even though the system might be 
physically 2D (like a 2D Fermi-Hubbard model), the computational method treats the system as 
a 1D chain by mapping the 2D lattice onto a 1D structure. This mapping is done by assigning a 
unique number to each site in the 2D grid. For example, a 2D square lattice can be transformed into a 1D chain:
\begin{center} 
$\begin{array}{ccc} 
1 & 2 & 3\\ 
4 & 5 & 6\\ 
7 & 8 & 9 
\end{array}$$\rightarrow$$1-2-3-4-5-6-7-8-9$ 
\par\end{center} 
Under this 1D representation: 

\begin{itemize} 
\item Sites that were horizontally adjacent in 2D (e.g., 1 and 2, 2 and 3) 
remain nearest neighbors in the 1D chain. 
\item However, vertically adjacent sites (e.g., 1 and 4, 2 and 5) become distant in the 1D mapping, introducing non-local interactions.
\end{itemize} 

This mapping creates a challenge, as non-local terms emerge when a 2D lattice is flattened into a 1D chain. When a 2D lattice is mapped to 1D (as 
done in traditional MPS/MPO approaches), nearest-neighbor interactions 
in the vertical direction of the 2D lattice, such as hopping terms between vertically adjacent sites, become non-local in the 1D representation.

\begin{itemize} 
\item This forces the tensor network to account for interactions between distant sites in the 1D chain.
\item These non-local interactions increase computational cost and slow convergence in methods like DMRG. 
\end{itemize} 

Thus, a basic solution to this is to find fermion-to-qubit maps that minimize the emergence of non-local terms. One possible path to explore is adopting in our algorithm the Cirac-Verstraete (CV) mapping (snake-like ordering). This method helps mitigate the issue of non-local terms that typically arise when mapping 2D systems, such as the 2D Fermi-Hubbard model, into a 1D representation. 

The CV mapping is implemented using a snake-like or zigzag ordering of the 2D lattice into a 1D chain, which is designed to reduce non-local interactions. In a traditional 1D mapping, the sites of a 2D grid are typically numbered in a simple row-by-row manner. However, the Cirac-Verstraete mapping improves upon this by zigzagging the numbering of sites so that both horizontal and vertical neighbors in the original 2D lattice remain as local neighbors in the 1D chain as much as possible. 

\begin{center} 
$\begin{array}{ccc} 
1 & 2 & 3\\ 
6 & 5 & 4\\ 
7 & 8 & 9 
\end{array}$$\rightarrow$$1-2-3-6-5-4-7-8-9$ 
\par\end{center} 

As a result, horizontal neighbors (e.g., 1 and 2, 2 and 3) remain adjacent in the 1D chain. Additionally, vertical neighbors (e.g., 1 and 6, 2 and 5) are now mapped closer together, reducing the number of long-range non-local interactions. This implementation within the MPO has several key advantages: 
\begin{itemize} 
\item Nearest-neighbor hopping terms in the Hamiltonian can now be represented using local or near-local MPO tensors. 
\item The reduction in long-range interactions improves the efficiency of tensor network contraction and simulation.
\end{itemize}  

Thus, a basic solution to this is to find fermion-to-qubit maps that minimize the emergence of non-local terms. One possible path to explore is adopting in our algorithm the Cirac-Verstraete (CV) mapping (snake-like ordering). This method helps mitigate the issue of non-local terms that typically arise when mapping 2D systems, such as the 2D Fermi-Hubbard model, into a 1D representation. 

The CV mapping is implemented using a snake-like or zigzag ordering of the 2D lattice into a 1D chain, which is designed to reduce non-local interactions. In a traditional 1D mapping, the sites of a 2D grid are typically numbered in a simple row-by-row manner. However, the Cirac-Verstraete mapping improves upon this by zigzagging the numbering of sites so that both horizontal and vertical neighbors in the original 2D lattice remain as local neighbors in the 1D chain as much as possible. 

\begin{center} 
$\begin{array}{ccc} 
1 & 2 & 3\\ 
6 & 5 & 4\\ 
7 & 8 & 9 
\end{array}$$\rightarrow$$1-2-3-6-5-4-7-8-9$ 
\par\end{center} 

As a result, horizontal neighbors (e.g., 1 and 2, 2 and 3) remain adjacent in the 1D chain. Additionally, vertical neighbors (e.g., 1 and 6, 2 and 5) are now mapped closer together, reducing the number of long-range non-local interactions. This implementation within the MPO has several key advantages:  

\begin{itemize} 
\item Nearest-neighbor hopping terms in the Hamiltonian can now be represented using local or near-local MPO tensors. 
\item The reduction in long-range interactions improves the efficiency of tensor network contraction and simulation.
\end{itemize}  

In summary, the CV mapping transforms the 2D lattice into a 1D chain in such a way that it 
preserves the nearest-neighbor relationship between both horizontal and vertical neighbors 
as much as possible. By doing so, it significantly reduces the number of non-local terms in the MPO representation, improving computational efficiency by mitigating the impact of long-range interactions.

It is worth noting that alternative mappings, such as those discussed in \cite{JW_1_Derby_2021, JW_2_Chiew_2023, JW_3_localfermionic}, can outperform the Cirac-Verstraete (CV) mapping in certain scenarios. However, in our tests using error-mitigated transcorrelated 
DMRG, we did not find any significant advantage in using more sophisticated mappings. 
This may be attributed to the fact that our implementation relies on a time-independent DMRG algorithm. 
More advanced mappings could potentially provide advantages in scenarios where alternative ansätze are used to initialize the MPO with projected wavefunctions.  

\subsection{DMRG Algorithm}

DMRG is a highly efficient classical algorithm for investigating the quantum properties of many-body systems. Originally developed to study the ground states of quantum spin chains, DMRG is particularly effective in one-dimensional (1D) systems, where it provides highly accurate results by reducing computational complexity through systematic truncation of the state space. DMRG is primarily used to explore ground-state energies, correlation functions, and low-lying excited states in condensed matter systems, offering a means to handle complex systems with many interacting particles.

DMRG operates by iteratively optimizing a wavefunction represented in MPS form. The MPS representation enables DMRG to focus on the most relevant quantum states while discarding those with minimal contributions to the ground-state properties. By capturing only the significant portions of the wavefunction, DMRG makes it feasible to study systems that would otherwise be intractable due to the exponential growth of the state space.

To further explore the seminal works of the traditional DMRG algorithm, the reader may want to check \cite{White1992, White1993, Ostlund1995, Dukelsky1998, Vidal2003}.

\subsubsection{Two-Site DMRG Algorithm}

The two-site DMRG algorithm provides an intuitive approach to finding the ground state of a 1D quantum system by sequentially adding sites and optimizing the MPS representation of the wavefunction. This version of DMRG is particularly useful for pedagogical purposes, as it offers a clear path for understanding the algorithm’s inner workings.

\begin{enumerate}
    \item \textbf{Wavefunction Representation with MPS}  
    The DMRG wavefunction is expressed as an MPS, where each site \( i \) is represented by a tensor \( A_i \). For a system with \( L \) sites, the wavefunction can be decomposed as:
    \[
    |\Psi\rangle = \sum_{\{n\}} A_1^{n_{1}} A_2^{n_{2}} \ldots A_L^{n_{L}} |n_{1} n_{2} \ldots n_{L}\rangle,
    \]
    where each tensor \( A_i^{n_{i}} \) is an \( m \times m \) matrix, and \( m \), the bond dimension, controls the size and accuracy of the representation. The integers \( n_{i} \) indicate the occupation number for each site. The MPS form enables efficient representation of the wavefunction in large-dimensional Hilbert spaces.

    \item \textbf{Density Matrix Truncation}  
    At each DMRG step, the algorithm computes the reduced density matrix for the block of sites currently being optimized. The eigenvalues of the density matrix indicate the importance of the states. By retaining only the states with the largest eigenvalues, DMRG systematically reduces the size of the Hilbert space while preserving the essential part of the wavefunction, ensuring efficient convergence to the ground state.

    \item \textbf{Optimization of the Ground State with the Davidson Algorithm}  
    DMRG minimizes the ground-state energy using the variational principle:
    \[
    E_0 = \min_{|\Psi\rangle} \frac{\langle \Psi|\hat{H}|\Psi\rangle}{\langle \Psi|\Psi\rangle},
    \]
    where \( \hat{H} \) is the Hamiltonian of the system, and \( E_0 \) is the ground-state energy. The optimization involves solving a large eigenvalue problem for each two-site subsystem, which is efficiently handled using the \textbf{Davidson algorithm}, an iterative solver designed for large sparse matrices.

    \item \textbf{Effective Hamiltonian and Local Optimization}  
    During the DMRG sweep, the Hamiltonian is reformulated as an effective Hamiltonian for the two optimized sites:
    \[
    H_{\text{eff}} \Psi_{\text{eff}} = E_{\text{eff}} \Psi_{\text{eff}},
    \]
    where \( H_{\text{eff}} \) is the effective Hamiltonian matrix for the pair of sites being optimized, and \( \Psi_{\text{eff}} \) is the effective ground-state wavefunction for these sites. The effective Hamiltonian \( H_{\text{eff}} \) is typically Hermitian, and solving this linear eigenvalue problem provides the energy \( E_{\text{eff}} \) and the updated state \( \Psi_{\text{eff}} \).
\end{enumerate}

\subsubsection{Davidson Algorithm for Eigenvalue Problems in DMRG}

The Davidson algorithm is a popular choice for solving large, sparse eigenvalue problems, particularly when the goal is to find only a few of the lowest eigenvalues and their corresponding eigenvectors, as in DMRG. The key advantage of the Davidson algorithm is that it utilizes the sparsity of the Hamiltonian matrix, avoiding the need to store or operate on the full matrix. The Davidson algorithm  consists of the following steps:
\begin{enumerate}
    \item \textbf{Initialization:} We start with an initial guess for the ground-state eigenvector \( |\Psi_0\rangle \), chosen randomly or based on prior knowledge of the system. This initial vector belongs to a subspace spanned by a basis \( \{|\phi_i\rangle\} \), where initially \( i = 1 \).

    \item \textbf{Subspace Expansion:} At each iteration, the algorithm projects the Hamiltonian \( \hat{H} \) into the current subspace spanned by the basis vectors \( \{|\phi_i\rangle\} \), forming a small projected Hamiltonian matrix \( H_{\text{sub}} \). The eigenvalues of \( H_{\text{sub}} \) are then computed, yielding an approximate ground-state energy \( E_0 \) and approximate eigenvector \( |\Psi\rangle \) in the subspace.

    \item \textbf{Residual Calculation:} The residual vector \( r \) is calculated as
    \begin{equation}
    r = \hat{H}|\Psi\rangle - E_0 |\Psi\rangle.
    \end{equation}
    This residual measures how well the approximate eigenvector satisfies the eigenvalue equation \( \hat{H}|\Psi\rangle = E_0 |\Psi\rangle \). A smaller residual indicates a closer approximation to the true eigenvector.

    \item \textbf{Correction Vector:} If the residual is not small enough (indicating that the subspace does not yet contain an accurate ground state), a correction vector \( |v\rangle \) is computed to expand the subspace. For symmetric matrices, the correction vector can be estimated by
    \begin{equation}
    |v\rangle \approx \frac{r}{\hat{H} - E_0 I},
    \end{equation}
    where \( I \) is the identity matrix. This equation is solved approximately, avoiding an exact inversion, and adding \( |v\rangle \) to the subspace expands the basis, improving the approximation of the ground state.

    \item \textbf{Iteration and Convergence:} Steps 2-4 are repeated iteratively, with each iteration expanding the subspace and refining the eigenvalue and eigenvector estimates. The algorithm converges when the residual \( r \) is sufficiently small, at which point the approximate eigenvalue \( E_0 \) and eigenvector \( |\Psi\rangle \) are considered accurate.
\end{enumerate}

\subsubsection{MPS Structure in DMRG}

DMRG’s strength lies in its use of MPS to represent quantum states. In an MPS, the wavefunction is expressed as a product of matrices, allowing efficient storage and manipulation. The outermost matrices are vectors of dimensions \( 1 \times m \) and \( m \times 1 \), respectively, while the inner matrices are of size \( m \times m \). The parameter \( m \), termed the bond dimension, governs the complexity and accuracy of the representation. A higher bond dimension allows for a more accurate representation of entanglement at the cost of increased computational resources.

\subsubsection{Time-Independent and Time-Dependent DMRG}

The above discussion focuses on time-independent DMRG, which is specifically designed for finding ground-state properties. However, DMRG can also be extended to handle time-dependent problems, where the goal is to simulate the real-time or imaginary-time evolution of a quantum system.

It has been extended to simulate time evolution through the time-dependent DMRG (tDMRG) method \cite{Daley2004,White2004}. This extension has enabled researchers to explore the dynamics of quantum states governed by both time-dependent and time-independent Hamiltonians. Here, we focus on the application of tDMRG to time-independent Hamiltonians, which has proven highly effective in capturing non-equilibrium and dynamical properties. Time evolution in tDMRG is governed by the Schr\"odinger equation:
\begin{equation}
    i \frac{\partial}{\partial t} |\psi(t)\rangle = H |\psi(t)\rangle,
\end{equation}
where $H$ is the time-independent Hamiltonian and $|\psi(t)\rangle$ is the time-evolved quantum state. The state is represented as MPS, which efficiently encodes the wavefunction when entanglement is limited. The dynamics are simulated numerically using schemes such as the Trotter-Suzuki decomposition or the time-dependent variational principle (TDVP). The Trotter-Suzuki approach decomposes the Hamiltonian into terms that act locally, applying their effects sequentially over small time steps, while the TDVP projects the dynamics onto the manifold of MPS states, allowing for more adaptive and often more accurate evolution.

Applications of tDMRG to time-independent Hamiltonians are diverse. A key use case is the study of quench dynamics, where the system is initialized in the ground state of one Hamiltonian and evolved under a different Hamiltonian after a sudden parameter change. This setup provides insights into relaxation processes, thermalization, and transport phenomena. For instance, current or heat transport can be investigated by preparing a system with an initial imbalance and observing how it evolves over time. Other applications include the propagation of excitations or domain walls in spin chains and the study of relaxation towards equilibrium in integrable and non-integrable systems.

The success of tDMRG arises from its ability to handle low-entanglement states effectively, which makes it ideal for studying short-to-intermediate time dynamics in one-dimensional systems. However, its primary limitation is the growth of entanglement entropy during time evolution, which increases the computational cost exponentially by requiring larger bond dimensions in the MPS representation. This limits the available timescales, particularly for systems with rapid entanglement growth or long-range interactions.

Despite these challenges, tDMRG offers several advantages. It provides high accuracy for systems with moderate entanglement and allows systematic control of numerical errors, such as those arising from Trotter decomposition or MPS truncation. Recent developments have further enhanced its applicability, such as using the TDVP method to mitigate Trotter errors and employing disentangling techniques or symmetries to slow entanglement growth.

In conclusion, tDMRG is a versatile and robust tool for simulating the dynamics of quantum systems governed by time-independent Hamiltonians. Its ability to capture non-equilibrium phenomena and relaxation dynamics has made it an essential method in computational quantum physics. While the entanglement barrier remains a significant limitation, ongoing advances in algorithms and computational strategies hold promise for extending its reach to longer times and more complex systems.

\subsubsection{Trotter Decomposition and Trotter Errors in t-DMRG}

To implement time evolution in DMRG, the time-evolution operator \( e^{-\tau \hat{H}} \) is typically decomposed using the \textbf{Trotter-Suzuki decomposition}. This method approximates the exponential of a sum of operators by breaking it into a product of exponentials of individual operators. However, this decomposition introduces \textbf{Trotter errors}, which arise from the non-commutativity of the Hamiltonian terms. These errors are controlled by the time step \( \Delta t \); smaller time steps reduce the error but increase computational cost. For accurate time-dependent DMRG simulations, a balance between the time step size and computational resources must be maintained. In systems requiring high precision, the cumulative Trotter error can become significant, potentially affecting the fidelity of the simulation, especially over long time evolutions.

\subsubsection{Limitations and Extensions of DMRG}

While DMRG excels in 1D systems and systems with low entanglement, it becomes less efficient in two-dimensional (2D) systems, where the entanglement grows more rapidly. In 2D, maintaining an accurate MPS representation requires a significantly larger bond dimension, making the method computationally challenging for large systems. Consequently, for 2D systems or to capture strong dynamic correlations, other approaches, such as \textbf{tensor network methods} or \textbf{projected entangled pair states (PEPS)}, may be required.

In summary, DMRG is an efficient classical algorithm designed to compute the ground-state properties of quantum many-body systems by iteratively optimizing a wavefunction in MPS form. The two-site DMRG algorithm reduces computational demands by focusing on a two-site subsystem at each optimization step. By discarding low-weight states in the density matrix, DMRG keeps the computation feasible while retaining the most relevant quantum states, achieving high accuracy in 1D systems. The Davidson algorithm serves as an efficient iterative eigenvalue solver, crucial for large-scale DMRG calculations. With extensions for time-dependent calculations, DMRG has become a fundamental tool in condensed matter physics and quantum chemistry for exploring static and dynamic properties of complex quantum systems. However, the Trotter errors in time evolution must be carefully controlled to ensure simulation accuracy, especially in long-time dynamics.

\section{\label{4} Numerical Experiments: EMTC-DMRG}

The numerical validation presented here focuses on the transcorrelated Fermi-Hubbard model, which serves as a rigorous testbed for our approach. Given that EMTC-DMRG is designed to handle transcorrelated fermionic systems, further case studies would provide similar results, as the key improvements—error mitigation and computational efficiency—are independent of the specific model choice. As was explained in the section \ref{3}, we will test our algorithm for the Fermi-Hubbard model under some regimes:

\begin{itemize}
\item \textbf{Weakly Correlated Regime (\( U/t \in [0, 2] \)):} In this regime, the kinetic energy dominates over the Coulomb repulsion. Electrons can move freely between sites, resulting in a metallic-like state. The electronic correlations are predominantly dynamic, and low-order perturbation theory is sufficient to describe the system accurately.
\item \textbf{Intermediately Correlated Regime (\( U/t \in [4, 8] \)):} Here, the kinetic energy and Coulomb repulsion become comparable. Significant electronic correlations emerge, and the system may transition toward a Mott insulating state, depending on the filling and other parameters. Both static and dynamic correlations coexist, making this regime challenging to address with traditional methods. Numerical techniques, such as the DMRG, are often required.
\item \textbf{Strongly Correlated Regime (\( U/t \in [10, 20] \)):} In this regime, the Coulomb repulsion dominates. Double occupancy of lattice sites is strongly suppressed, leading to Mott insulating behavior. The electron mobility is reduced, and the system's dynamics are governed by superexchange interactions, typically scaling as \( t^2/U \). Effective spin models, such as the Heisenberg model, often describe the system well.
\end{itemize}

We applied our EMTC-DMRG, explained in Section \ref{2}, to the two-dimensional Fermi-Hubbard model, considering periodic boundary conditions, different lattice sizes, fillings, and interaction strengths. In our approach, as described in Section \ref{3}, we use time-independent DMRG (TI-DMRG) and map the 2D lattices to 1D using a snake-like model. We combined this with an analytical formulation of the transcorrelated transformation for the target system. Furthermore, our Hamiltonian is described in real space throughout this section, while the TC-DMRG method, with which we compare in this section, is described using a k-space Hamiltonian.

The numerical experiments presented here were conducted using a Python-based code. Our Hamiltonian was implemented using the \texttt{ OpenFermion} library, with an interface developed for the \texttt{Renormalizer} library, as proposed by \cite{bipartite}. The exact reference values used were obtained from \cite{ref_ground_state2, ref_ground_state1}. The outcomes presented on the tables, the Jastrow factor labeled as $J_{our}$ would be related to EMTC-DMRG and labeled as $J_{other}$ is related to the outcomes of the TC-DMRG presented in \cite{baiardi}. Our simulations were performed using an NVIDIA GeForce RTX 3050 Laptop GPU. 

We compared our EMTC-DMRG with the TC-DMRG obtained by \cite{baiardi}. In one of the steps of their formulation, they also used the two-site TI-DMRG variant to try to avoid convergence into local minima of the energy functional, as can happen in the one-site variant. All values of the ground-state energies are represented per site and in units of the hopping parameter \( t \). 

We first analyzed our EMTC-DMRG for a lattice \( 3 \times 3 \) with \( U=8 \), \( t=1 \), \( N_{\alpha}=4 \), and \( N_{\beta}=4 \). We report in Table \ref{tab:ground-state-energy} the results of the computed energies for the bond dimensions \( m=70, 90, 100, 200, 300 \). The results obtained by the concurrent methodology were reported for the bond dimensions \( m=100, 200, 300 \). We observed that, even though our methodology has the Hamiltonian described in real space, we already realized faster convergence to the ground state when we turned on the transcorrelator factor \( J \). Its negative value corresponds to the right eigenvector of the Hamiltonian, which is reported in the literature as being responsible for providing a more compact wavefunction.

In the concurrent method, the transcorrelation does not make a difference for \( m=100 \) and \( m=200 \), while for \( m=300 \), it converges to the true value obtained with exact diagonalization. However, \( J=0 \) also converged at this same bond dimension, which means that for this situation, the transcorrelation does not improve the convergence. We analyzed the same values of the initial bond dimensions reported in the TC-DMRG method and also added an analysis for lower values, including \( m=70 \),\( m=90 \) and \( m=100 \).

Surprisingly, we ran some tests as shown in Table \ref{tab:ground-state-energy}, where we can see that the results from a bond dimension of $m=70$ and $J=-0.15$ reach values close to the true ground state $-0.0012$. The results obtained with our method converged within $5$ to $7$ sweeps. It is important to realize that in our approach, as we do not use imaginary time evolution DMRG in one of the steps or the entire procedure, we have the breaking  of the  variational principle since we are dealing with a non-hermitian.

So, that indicates that we are reducing physical resources since a significant reduction of the value of the bond dimension. Also, in our approach, the transcorrelation makes a difference and helps to the faster convergence regarding the original model. We see that if we keep increasing the transcorrelation parameter $J$, the ground state values keep going below the true value, but in this case still is physically acceptable, and we chose the value of the bond dimension closer to the exact value.

\begin{figure}[H]
    \centering
    \includegraphics[width=0.75\textwidth]{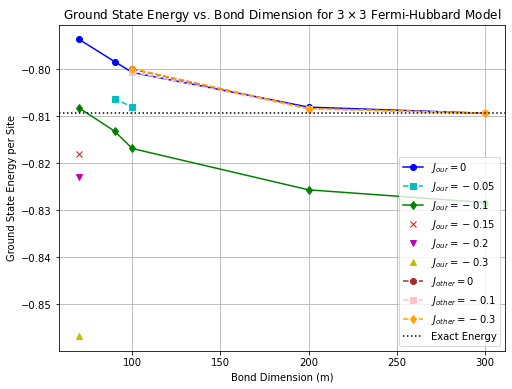}
    \caption{Ground state energy vs. bond dimension for the \(3 \times 3\) Fermi-Hubbard model, corresponding to Table \ref{tab:ground-state-energy}.}
    \label{fig:ground-state-energy-3x3}
\end{figure}

\begin{figure}[H]
    \centering
    \includegraphics[width=0.75\textwidth]{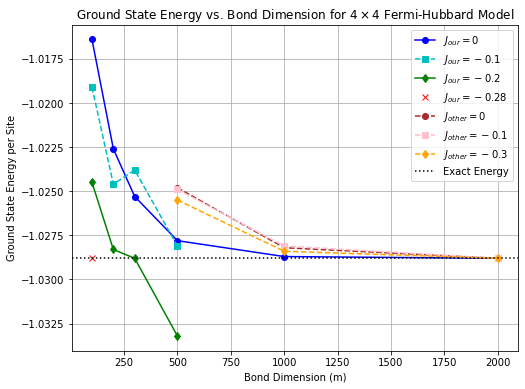}
    \caption{Ground state energy vs. bond dimension for the \(4 \times 4\) Fermi-Hubbard model, corresponding to Table \ref{tab:ground-state-energy-4x4}.}
    \label{fig:ground-state-energy-4x4}
\end{figure}

\begin{figure}[H]
    \centering
    \includegraphics[width=0.75\textwidth]{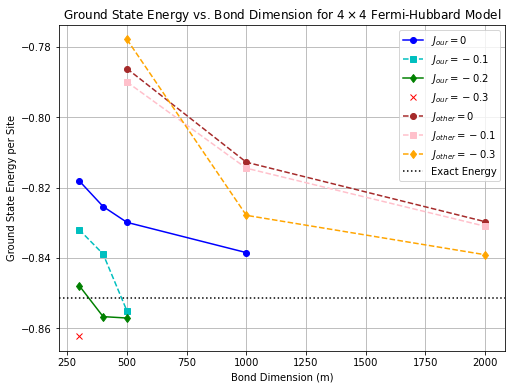}
    \caption{Ground state energy vs. bond dimension for the \(4 \times 4\) Fermi-Hubbard model at \( U/t=4 \), corresponding to Table \ref{tab:3}.}
    \label{fig:ground-state-energy-4x4-U4}
\end{figure}

To enhance readability, we provide visual representations of these numerical results in Figures \ref{fig:ground-state-energy-3x3},\ref{fig:ground-state-energy-4x4}, and \ref{fig:ground-state-energy-4x4-U4} which illustrate the trends observed in Tables \ref{tab:ground-state-energy},\ref{tab:ground-state-energy-4x4}, and \ref{tab:3}.

\begin{table}[H]
\centering
\renewcommand{\arraystretch}{1.2} 
\setlength{\tabcolsep}{4pt} 
\scalebox{0.65}{ 
\begin{tabular}{|c|c|c|c|c|c|c|c|c|c|}
\hline
$m$ & $J_{our}=0$ & $J_{our}=-0.05$ & $J_{our}=-0.1$ & $J_{our}=-0.15$ & $J_{our}=-0.2$ & $J_{our}=-0.3$ & $J_{other}=0$ & $J_{other}=-0.1$ & $J_{other}=-0.3$ \\ \hline \hline
$70$ & -0.7937 & - & -0.8083 & -0.8181 & -0.8230 & -0.8568 & - & - & - \\ \hline
$90$ & -0.7984 & -0.8063 & -0.8132 & - & - & - & - & - & - \\ \hline
$100$ & -0.8007 & -0.8081 & -0.8169 & - & - & - & -0.8000 & -0.8006 & -0.7999 \\ \hline
$200$ & -0.8081 & - & -0.8257 & - & - & - & -0.8084 & -0.8085 & -0.8084 \\ \hline
$300$ & -0.8094 & -  & -0.8284 & - & - & - & -0.8094 & -0.8094 & -0.8094 \\ \hline
\end{tabular}
} 
\caption{Ground state energy per site for $3x3$ Fermi-Hubbard model, pbc, for $U=8$, $t=1$, $N_{\alpha}=4$, $N_{\beta}=4$, considering varying bond dimension $m$. The reference energy obtained by exact diagonalization is $-0.8094$. }
\label{tab:ground-state-energy}
\end{table}

In Table \ref{tab:ground-state-energy-4x4}, we increase the lattice size to \( 4 \times 4 \), using the same filling and interaction strength parameters. Here, we are now comparing with the TC-DMRG formulation in k-space, while our methodology remains in real space. As reported, we observe that the concurrent method achieves convergence of the ground-state energies at a bond dimension of \( m=2000 \), and the transcorrelated term does not provide any improvement compared to the original Fermi-Hubbard Hamiltonian. In contrast, as we demonstrate, with EMTC-DMRG, we achieve convergence to the true ground-state energy value with a bond dimension of \( m=100 \) and \( J=-0.28 \). This result highlights a drastic reduction in the initial bond dimension required for the execution of the algorithm using our methodology.

\begin{table}[H]
\centering
\renewcommand{\arraystretch}{1.2} 
\setlength{\tabcolsep}{4pt} 
\scalebox{0.65}{ 
\begin{tabular}{|c|c|c|c|c|c|c|c|}
\hline
$m$ & $J_{our}=0$ & $J_{our}=-0.1$ & $J_{our}=-0.2$ & $J_{our}=-0.28$ & $J_{other}=0$ & $J_{other}=-0.1$ & $J_{other}=-0.3$ \\ \hline \hline
$100$ & -1.0164 & -1.0191 & -1.0245 & -1.0288 & - & - & - \\ \hline
$200$ & -1.0226 & -1.0246 & -1.0283 & - & - & - & - \\ \hline
$300$ & -1.0253 & -1.0238 & -1.0288 & - & - & - & - \\ \hline
$500$ & -1.0278 & -1.0281 & -1.0332 & - & -1.0248 & -1.0249 & -1.0255 \\ \hline
$1000$ & -1.0287 & - & - & - & -1.0282 & -1.0281 & -1.0284 \\ \hline
$2000$ & -1.0288 & - & - & - & -1.0288 & -1.0288 & -1.0288 \\ \hline
\end{tabular}
} 
\caption{Ground state energy per site for $4x4$ Fermi-Hubbard model, pbc, for $U=8$, $t=1$, $N_{\alpha}=4$, $N_{\beta}=4$, considering varying bond dimension $m$. The reference energy obtained by exact diagonalization is $-1.0288$. The values reported in the literature were obtained in k space ($J_{other}$).}
\label{tab:ground-state-energy-4x4}
\end{table}

In Table~\ref{tab:3}, we report results for the Fermi-Hubbard model with \( 4 \times 4 \), \( U/t=4 \), \( N_{\alpha}=8 \), and \( N_{\beta}=8 \). We continue comparing our EMTC-DMRG with TC-DMRG. Under these conditions, even in k-space with a bond dimension of \( m=2000 \), the TC-DMRG method could not achieve convergence to the true energy value. In contrast, with our methodology, we achieve convergence using a bond dimension of \( m=300 \) and a correlation factor of approximately \( J \sim -0.25 \), demonstrating a clear advantage of our approach. 

The number of sweeps required in our simulations involving the transcorrelation term is just $4$ for most cases. The higher bond dimensions observed in the previous two cases can be attributed to the breakdown of the area law for long-range Hamiltonians, as discussed in \cite{baiardi}. However, in this case, with an optimal MPO construction free from numerical errors and time-independent DMRG (free of Trotter errors), we obtain satisfactory results with significantly smaller bond dimensions.

Unlike the works of \cite{baiardi,dobrautz}, the transcorrelated ansatz produces a more compact MPS even in real space, when the EMTC-DMRG is applied to the two-dimensional Fermi-Hubbard Hamiltonian.
\begin{table}[H]
\centering
\renewcommand{\arraystretch}{1.2} 
\setlength{\tabcolsep}{4pt} 
\scalebox{0.65}{ 
\begin{tabular}{|c|c|c|c|c|c|c|c|}
\hline
$m$ & $J_{our}=0$ & $J_{our}=-0.1$ & $J_{our}=-0.2$ & $J_{our}=-0.3$ & $J_{other}=0$ & $J_{other}=-0.1$ & $J_{other}=-0.3$ \\ \hline \hline
$300$  & -0.8181 & -0.8320 & -0.8479 & -0.8622 & - & - & - \\ \hline
$400$  & -0.8254 & -0.8389 & -0.8567 & - & - & - & - \\ \hline
$500$  & -0.8299 & -0.8551 & -0.8571 & - & -0.7862 & -0.7900 & -0.7779 \\ \hline
$1000$ & -0.8385 & - & - & - & -0.8128 & -0.8145 & -0.8279 \\ \hline
$2000$ & - & - & - & - & -0.8297 & -0.8310 & -0.8391 \\ \hline
\end{tabular}
} 
\caption{Ground state energy per site for $4x4$ Fermi-Hubbard model (tc DMRG) for $U=4$, $t=1$, $N_{\alpha}=8$, $N_{\beta}=8$, considering varying bond dimension $m$. The reference energy obtained by exact diagonalization is $-0.8514$.}
\label{tab:3}
\end{table}

Here we just analyze some other conditions that were not explored in the TC DMRG method and we compare with the exact value. 
\begin{table}[H]
\centering
\renewcommand{\arraystretch}{1.2} 
\setlength{\tabcolsep}{4pt} 
\scalebox{1}{ 
\begin{tabular}{|c|c|c|c|}
\hline
$m$ & $J_{our}=0$ & $J_{our}=-0.1$ & $J_{our}=-0.3$ \\ \hline \hline
$100$ & -0.7002 & -0.7003 & -0.7008 \\ \hline
\end{tabular}
} 
\caption{Ground state energy per site for $4x4$ Fermi-Hubbard model (tc DMRG) for $U=12$, $t=1$, $N_{\alpha}=2$, $N_{\beta}=2$, considering varying bond dimension $m$. The reference energy obtained by exact diagonalization is $-0.7003$.}
\label{tab:4}
\end{table}
 In table \ref{tab:4}, with the very strongly correlated regime, but very low filling, we have the initial bond dimension requested to convergence of the code being low. In this case, the Jastrow factor does not make too much difference from the original Hamiltonian, which is just around $0.0001$.
\begin{table}[H]
\centering
\renewcommand{\arraystretch}{1.2} 
\setlength{\tabcolsep}{4pt} 
\scalebox{1}{ 
\begin{tabular}{|c|c|c|c|}
\hline
$m$ & $J_{our}=0$ & $J_{our}=-0.1$ & $J_{our}=-0.4$ \\ \hline \hline
$100$ & -0.3237 & -0.3320 & -0.3678 \\ \hline
$300$ & -0.3613 & -0.3745 & - \\ \hline
\end{tabular}
} 
\caption{Ground state energy per site for $4x4$ Fermi-Hubbard model (tc DMRG) for $U=12$, $t=1$, $N_{\alpha}=8$, $N_{\beta}=8$, considering varying bond dimension $m$. The reference energy obtained by exact diagonalization is $-0.3745$.}
\label{tab:5}
\end{table}
In the table \ref{tab:5}, in the same system, when we increase the filling, we also need to increase the initial bond dimension, and here we can see the Jastrow factor $J=-0.1$ making a difference and helping to convergence faster than its original Hamiltonian.
\begin{table}[H]
\centering
\renewcommand{\arraystretch}{1.2} 
\setlength{\tabcolsep}{4pt} 
\scalebox{1}{ 
\begin{tabular}{|c|c|c|c|}
\hline
$m$ & $J_{our}=0$ & $J_{our}=-0.1$ & $J_{our}=-0.4$ \\ \hline \hline
$100$ & -0.9004 & -0.8944 & -0.8825 \\ \hline
$300$ & -0.9060 & -0.9061 & -0.9072 \\ \hline
\end{tabular}
} 
\caption{Ground state energy per site for $4x4$ Fermi-Hubbard model (tc DMRG) for $U=12$, $t=1$, $N_{\alpha}=3$, $N_{\beta}=3$, considering varying bond dimension $m$. The reference energy obtained by exact diagonalization is $-0.9061$. Simulations obtained by DMRG two sites (sweep algorithm also known as time-independent DMRG) in the real basis.}
\label{tab:6}
\end{table}
Again, in Table \ref{tab:6}, when we reduce the filling, we can barely observe the impact that a non-zero Jastrow factor can produce. This suggests that, in this regime, it is more efficient to apply the Jastrow factor for higher fillings. However, as the lattice size increases, it becomes necessary to provide the k-momentum formulation of the TC-FH Hamiltonian. It is important to highlight the limitations of our methodology in real space, which begin to emerge as the lattice size increases. Specifically, from a \( 6 \times 6 \) lattice onward, under the conditions reported in \cite{baiardi}, our approach no longer offers an advantage in reducing the initial bond dimension for these systems, for instance:
\begin{itemize}
    \item $U/t=4$, $N_{\alpha}=N_{\beta}=12$;
    \item $U/t=2$, $N_{\alpha}=N_{\beta}=18$;
    \item $U/t=4$, $N_{\alpha}=N_{\beta}=18$;
    \item $U/t=4$, $N_{\alpha}=N_{\beta}=18$;
    
\end{itemize}
So, in this case, we suggest using the k-space Hamiltonian as described in Eq.\ref{kspace} instead of the real one. Probably with this change our method would be competitive for this lattice size. 

We also attempted to carry out an imaginary time evolution using EMTC-DMRG; however, the computational cost was prohibitively high and unsuitable for execution on a laptop in most cases. Additionally, we tried optimizing the MPS using time-independent DMRG and then applying the imaginary time evolution, as described in \cite{baiardi}. However, this approach did not yield better results regarding the convergence of the method, and it proved to be computationally more expensive than the version we present in this work.

Another important point to highlight is that our classical variational algorithm, which employs time-independent DMRG (TI-DMRG), produces results that are independent of the fermion-to-qubit mapping. We tested alternative mapping methods beyond snake-like mapping \cite{Verstraete_2005}. For example, more sophisticated approaches, such as those described in \cite{JW_1_Derby_2021, JW_2_Chiew_2023, JW_3_localfermionic}, did not result in improved performance for EMTC-DMRG. However, they might lead to enhancements in versions utilizing the imaginary time-evolution DMRG algorithm.

To summarize the differences between our EMTC-DMRG and existing transcorrelated DMRG approaches, we provide a comparative in Table \ref{tab:TC-DMRG}. This table highlights the key methodological distinctions, computational trade-offs, and advantages of each approach. Notably, our method improves numerical stability through symbolic optimization and achieves faster convergence without requiring imaginary-time evolution, unlike previous TC-DMRG formulations.

\begin{table*}[ht]
    \centering
    \renewcommand{\arraystretch}{1.5} 
    \setlength{\tabcolsep}{8pt} 
    \resizebox{\textwidth}{!}{
    \begin{tabular}{|>{\columncolor{yellow!20}}l|>{\columncolor{purple!15}}m{5cm}|>{\columncolor{blue!15}}m{5cm}|>{\columncolor{green!15}}m{5cm}|}
        \hline
        \textbf{Algorithm} & \textbf{EMTC-DMRG} & \textbf{TC-DMRG \cite{baiardi}} & \textbf{TC-DMRG \cite{TC_DMRG_Alavi_liao2023density}} \\
        \hline
        \textbf{Main Approach} & Symbolic optimization; TI-DMRG applied to a TC Hamiltonian described in real space to mitigated errors. &  Uses TI-DMRG and iTD- DMRG and  applied to TC Hamiltonian  described in k-space. & Time-independent DMRG with a generalized Davidson solver for non-Hermitian Hamiltonians \\
        \hline
        \textbf{Key Features} & Optimal exact MPO; Analytical TC-FH Hamiltonian; Standard Davidson solver & Exact MPO; Analytical TC- FH Hamiltonian; Standard Davidson solver. & Compressed MPO; Approximated TC Hamiltonian formulation; Generalized Davison solver.  \\
        \hline
        \textbf{Challenges Addressed} & Reducing long-range interactions; Show that TC helps to improve the 2D DMRG performance to compute ground state and in some cases outperforms \cite{baiardi}. & Reducing long-range interactions; Show that TC helps to improve the 2D DMRG performance to compute ground state. &  Shows that they TC formulation improves the efficiency and accuracy in computingt ground/excited state of some molecules. \\
        \hline
        \textbf{Target Systems} & 2D Fermi-Hubbard model with pbc. & 2D Fermi-Hubbard model with pbc. & Molecular systems. \\
        \hline
        \textbf{Computational Efficiency} & Significant reduction of the resources like bond dimension $m$ in relation to \cite{baiardi}.  & Significant reduction of the resources like bond dimension $m$ in relation to TI-DMRG. & Accelerated basis set convergence for molecular systems. \\
        \hline
        \textbf{Variational Principle} & Does not apply here. & Applies here due the use of imaginary time evolution. & Applies here due the use of Generalized Davidson solver for non-Hermitian Hamiltonian.\\
        \hline
        \textbf{Limitations} & Not optimal for very large lattices without a k-space Hamiltonian &  Requires imaginary-time evolution, and k-space formulation of the Hamiltonian even for modest size lattices. & Requires Generalized Davidson solver; Only suitable for molecular systems. \\
        \hline
    \end{tabular}
    }
    \caption{Comparison of Transcorrelated DMRG Methods.}
    \label{tab:TC-DMRG}
\end{table*}

Unlike previous TC-DMRG formulations, our method eliminates the need for imaginary-time evolution, reducing computational overhead while achieving comparable accuracy. This is reflected in Table \ref{tab:TC-DMRG}.

\section{\label{5} Discussions}

In this section, we apply the exact MPO method proposed by Jiajun et al. as described in the section \ref{3} to showcase their effect on the analytical Fermi-Hubbard Transcorrelated Hamiltonian which we are working with. 
The transcorrelated Fermi-Hubbard model in 2D for its analytical form is given by:
\begin{equation}
\begin{aligned}
\bar{H}=&-t\sum_{\langle i,j \rangle}a_{i,\sigma}^{\dagger}a_{j,\sigma}+U\sum_{l}n_{l,\uparrow}n_{l,\downarrow}\\
&-t\sum_{\langle i,j \rangle, \sigma}a_{i,\sigma}^{\dagger}a_{j,\sigma}\left\{(e^{J}-1)n_{j,\bar{\sigma}}+(e^{-J}-1)n_{i,\bar{\sigma}}-2\left[\cosh(J)-1\right]n_{i,\bar{\sigma}}n_{j,\bar{\sigma}}\right\}.
\end{aligned}
\end{equation}
With the goal to simplify the equations, we begin with the MPO decomposition for the Fermi-Hubbard term and later we will proceed with the extra terms which carry out the transcorrelated factor. By the end, we will sum the results obtained by both terms. So, considering the Fermi-Hubbard model on a 2D lattice. As demonstrated in previous section for the ab initio case, this process involves dividing the Hamiltonian into different parts and calculating their respective bond dimensions, which are key for efficient computational simulation using DMRG.
\begin{equation}
    H = -t \sum_{\langle i,j \rangle, \sigma} a_{i,\sigma}^\dagger a_{j,\sigma} + U \sum_{l} n_{l,\uparrow} n_{l,\downarrow}.
\end{equation}
where \(t\) is the hopping parameter, \(U\) is the on-site Coulomb repulsion, and \(\langle i, j \rangle\) represents nearest neighbors in the 2D lattice. The operators \(a_{i, \sigma}^\dagger\) and \(a_{i, \sigma}\) are the creation and annihilation operators for electrons of spin \(\sigma\) at site \(i\). The number operator is \(n_{i,\sigma} = a_{i,\sigma}^\dagger a_{i,\sigma}\).
The decomposition of the Hamiltonian is divided in three components:
\begin{enumerate}
\item \textbf{Intra-block terms (\(H_1\))}, which describe interactions within either the left or right block of the DMRG partition. 

\item \textbf{Inter-block two-electron terms (\(H_2\))}, which represent two-electron interactions coupling between the left and right blocks. 

\item \textbf{Inter-block three-electron terms (\(H_3\))}, which involve three operators from one block and one operator from the other. 
\end{enumerate}

The goal is to derive the bond dimensions for each component of the Hamiltonian. These bond dimensions determine the computational complexity of representing the MPO in the DMRG algorithm. The intra-block Hamiltonian describes all interactions within either the left or right block:
\begin{equation}
\begin{aligned}
    H_1 &= -t \sum_{\langle i_L, j_L \rangle, \sigma} a_{i_L, \sigma}^\dagger a_{j_L, \sigma} + U \sum_{i_L} n_{i_L, \uparrow} n_{i_L, \downarrow} \\
    &\quad -t \sum_{\langle i_R, j_R \rangle, \sigma} a_{i_R, \sigma}^\dagger a_{j_R, \sigma} + U \sum_{i_R} n_{i_R, \uparrow} n_{i_R, \downarrow}
\end{aligned}
\end{equation}
Here, \(i_L, j_L\) represent indices of the left block and \(i_R, j_R\) represent indices of the right block. Since the interactions occur only within individual blocks, the MPO bond dimension is minimal. The MPO only needs to represent two possible situations: either an identity operator or an intra-block Hamiltonian term. Thus, the bond dimension for \(H_1\) is:
\begin{equation}
    M_{O,1} = 4
\end{equation}
This value arises due to the hopping and interaction terms occurring both horizontally and vertically within the 2D lattice.
The inter-block two-electron Hamiltonian (\(H_2\)) involves terms that couple the left and right blocks across the partition boundary:
\begin{equation}
\begin{aligned}
    H_2 &= -t \sum_{\langle i_L, j_R \rangle, \sigma} a_{i_L, \sigma}^\dagger a_{j_R, \sigma} + \sum_{\langle i_R, j_L \rangle, \sigma} -t a_{i_R, \sigma}^\dagger a_{j_L, \sigma} \\
    &\quad + U \sum_{i_L, j_R} a_{i_L, \uparrow}^\dagger a_{i_L, \uparrow} a_{j_R, \downarrow}^\dagger a_{j_R, \downarrow}
\end{aligned}
\end{equation}
These terms represent hopping interactions across the boundary between the left and right blocks, as well as inter-block on-site Coulomb interactions. To represent these terms in an MPO efficiently, a complementary operator technique is employed. The complementary operator aggregates multiple hopping terms within a block into a single operator, effectively reducing the number of independent matrix elements needed for representation. For example, the complementary operator \(\hat{P}_{ij}\) is defined as:
\begin{equation}
    \hat{P}_{ij} = \sum_{p, r} -t a_{i_L, \sigma}^\dagger a_{j_R, \sigma}
\end{equation}
which allows multiple terms to be grouped and represented compactly. The bond dimension for \(H_2\) is:
\begin{equation}
    M_{O,2} = \min(n_L^2, n_R^2) + 2 \min\left( \frac{n_L (n_L - 1)}{2}, \frac{n_R (n_R - 1)}{2} \right)
\end{equation}
The first term \(\min(n_L^2, n_R^2)\) represents the number of ways two operators can interact across the partition, and the second term accounts for the possible internal pairings within each block. The inter-block three-electron Hamiltonian (\(H_3\)) involves interactions with three operators in one block and one operator from the other:
\begin{widetext}
\begin{equation}
\begin{aligned}
    H_3 = &\sum_{i_L} a_{i_L, \sigma}^\dagger \left( \sum_{j_R} -t a_{j_R, \sigma} + \sum_{k_R, l_R} U a_{k_R,\uparrow}^\dagger a_{k_R,\uparrow} a_{l_R, \sigma} \right) \\
    &+ \sum_{j_R} a_{j_R, \sigma} \left( \sum_{i_L} -t a_{i_L, \sigma}^\dagger + \sum_{k_L, l_L} U a_{k_L, \uparrow}^\dagger a_{k_L, \uparrow} a_{l_L, \sigma}^\dagger \right) \\
    &+ \sum_{k_R} \left( \sum_{l_L} -U a_{l_L, \downarrow}^\dagger a_{l_L, \downarrow} a_{k_R, \sigma} + \sum_{i_L, j_L} t a_{i_L, \sigma}^\dagger a_{j_L, \sigma} \right)
\end{aligned}
\end{equation}
\end{widetext}
This expanded form accounts for all possible interactions between three operators in one block and a single operator from the other block. Due to the large number of potential combinations, the use of complementary operators is further extended to aggregate interactions. For example, a complementary operator such as:
\begin{equation}
    \hat{Q}_{ijkl} = \sum_{p, q, r} U a_{p_L, \uparrow}^\dagger a_{q_L, \downarrow} a_{r_R, \sigma}
\end{equation}
is used to reduce the number of terms that need representation in the MPO. The bond dimension for \(H_3\) is then given by:
\begin{equation}
    M_{O,3} = 2\min\left( \frac{n_L^2 (n_L - 1)}{2}, n_R \right) + 2 \min\left( n_L, \frac{n_R^2 (n_R - 1)}{2} \right)
\end{equation}
This formula reflects the number of combinations of three operators from one block interacting with an operator from the other block. The total bond dimension for the entire MPO, \(M_O\), is the sum of the bond dimensions from \(H_1\), \(H_2\), and \(H_3\):
\begin{equation}
    M_{O,\text{max}} = M_{O,1} + M_{O,2} + M_{O,3}
\end{equation}
To simplify this expression for the case where the left and right blocks are of equal size (\(n_L = n_R = N/2\)), we get:
\begin{equation}
    M_{O,\text{max}} = 2 \left( \frac{N}{2} \right)^2 + 3 \left( \frac{N}{2} \right) + 4
\end{equation}
Thus, the total bond dimension grows quadratically with the number of orbitals, but complementary operators help to manage this growth, ensuring that the computational complexity remains feasible for practical DMRG calculations. The introduction of complementary operators aggregates multiple terms, thus reducing the number of independent matrix elements needed for the MPO representation. Let's extend our decomposition of the Hamiltonian to include the additional transcorrelated term. The given Hamiltonian for the transcorrelated Fermi-Hubbard model is:
\begin{equation}
\begin{aligned}
\bar{H} = &-t \sum_{\langle i,j \rangle} a_{i,\sigma}^{\dagger} a_{j,\sigma} + U \sum_{l} n_{l,\uparrow} n_{l,\downarrow} \\
&- t \sum_{\langle i,j \rangle, \sigma} a_{i,\sigma}^{\dagger} a_{j,\sigma} \left\{ (e^{J} - 1)n_{j,\bar{\sigma}} + (e^{-J} - 1)n_{i,\bar{\sigma}} - 2 \left[ \cosh(J) - 1 \right] n_{i,\bar{\sigma}} n_{j,\bar{\sigma}} \right\}
\end{aligned}
\end{equation}
This term introduces additional complexity involving two-body and three-body interactions, resulting from the transcorrelated transformation parameter $J$. Let\'s decompose this extended Hamiltonian into $H_1$, $H_2$, and $H_3$ and provide a detailed explanation of the MPO structure and complementary operators for the added terms. We have the additional transcorrelated interaction term:
\begin{equation}
-t\sum_{\langle i,j \rangle, \sigma}a_{i,\sigma}^{\dagger}a_{j,\sigma}\left\{(e^{J}-1)n_{j,\bar{\sigma}}+(e^{-J}-1)n_{i,\bar{\sigma}}-2\left[\cosh(J)-1\right]n_{i,\bar{\sigma}}n_{j,\bar{\sigma}}\right\}
\end{equation}
This term includes two-body and three-body interactions. We\'ll decompose these terms as follows:

The intra-block Hamiltonian $H_1$ for the transcorrelated term represents interactions that occur entirely within the left or right block. This can be expressed as:
\begin{equation}
\begin{aligned}
    \hat{H}_1 = &-t \sum_{\langle i_{L}, j_{L} \rangle, \sigma} a_{i_{L}, \sigma}^{\dagger} a_{j_{L}, \sigma} \left\{ (e^{J} - 1) n_{j_{L}, \bar{\sigma}} + (e^{-J} - 1) n_{i_{L}, \bar{\sigma}} \right\} \\
    &-2t \left[ \cosh(J) - 1 \right] \sum_{\langle i_{L}, j_{L} \rangle, \sigma} a_{i_{L}, \sigma}^{\dagger} a_{j_{L}, \sigma} n_{i_{L}, \bar{\sigma}} n_{j_{L}, \bar{\sigma}} \\
    &-t \sum_{\langle i_{R}, j_{R} \rangle, \sigma} a_{i_{R}, \sigma}^{\dagger} a_{j_{R}, \sigma} \left\{ (e^{J} - 1) n_{j_{R}, \bar{\sigma}} + (e^{-J} - 1) n_{i_{R}, \bar{\sigma}} \right\} \\
    &-2t \left[ \cosh(J) - 1 \right] \sum_{\langle i_{R}, j_{R} \rangle, \sigma} a_{i_{R}, \sigma}^{\dagger} a_{j_{R}, \sigma} n_{i_{R}, \bar{\sigma}} n_{j_{R}, \bar{\sigma}}
\end{aligned}
\end{equation}
Where  $i_L, j_L$ represent lattice sites within the left block, while $i_R, j_R$ represent sites within the right block. These terms involve hopping between nearest neighbors within each block. Since these interactions occur entirely within the left or right block, the bond dimension remains the same as for the original Hubbard model
\begin{equation}
M_{O,1} = 4
\end{equation}
The inter-block Hamiltonian $H_2$ involves two-body terms that couple the left and right blocks. The two-body interactions involve creation and annihilation operators from different blocks:
\begin{equation}
\begin{aligned}
    \hat{H}_2 = &-t \sum_{\langle i_L, j_R \rangle, \sigma} a_{i_L, \sigma}^{\dagger} a_{j_R, \sigma} \left\{ (e^{J} - 1) n_{j_R, \bar{\sigma}} + (e^{-J} - 1) n_{i_L, \bar{\sigma}} \right\} \\
    &- 2t \left[ \cosh(J) - 1 \right] \sum_{\langle i_L, j_R \rangle, \sigma} a_{i_L, \sigma}^{\dagger} a_{j_R, \sigma} n_{i_L, \bar{\sigma}} n_{j_R, \bar{\sigma}}.
\end{aligned}
\end{equation}
Where $i_L$ and $j_R$ denote lattice sites from the left and right blocks, respectively, and $\langle i_L, j_R \rangle$ represents nearest-neighbor interactions across the boundary.
To reduce the complexity of the MPO representation for $H_2$, we employ a complementary operator technique. For instance:
\begin{equation}
\hat{P}_{ij} = \sum_{k_L, l_R} -t \left\{ (e^{J} - 1) a_{k_L, \sigma}^\dagger n_{l_R, \bar{\sigma}} + (e^{-J} - 1) a_{k_L, \sigma}^\dagger n_{k_L, \bar{\sigma}} \right\}.
\end{equation}
This complementary operator aggregates multiple two-body interactions between the blocks, reducing the number of independent terms that need to be stored in the MPO representation. The bond dimension for $H_2$ then becomes:
\begin{equation}
M_{O,2} = \min(n_L^2, n_R^2) + 2 \min\left(\frac{n_L(n_L-1)}{2}, \frac{n_R(n_R-1)}{2}\right).
\end{equation}
Where the first term represents the number of ways to choose the inter-block two-electron interaction. The second term accounts for the interactions between sites within the left and right blocks. The $H_3$ term involves more complex interactions, including three-body terms that couple three operators in one block with a single operator from the other block. This decomposition can be expressed as:
\begin{equation}
\begin{aligned}
    \hat{H}_3 = &-2t \left[ \cosh(J) - 1 \right] \sum_{\langle i_L, j_R \rangle, \sigma} a_{i_L, \sigma}^{\dagger} a_{j_R, \sigma} n_{i_L, \bar{\sigma}} n_{j_R, \bar{\sigma}} \\
    &+ \sum_{i_L} a_{i_L, \sigma}^{\dagger} \left( \sum_{j_R} -t (e^J - 1) n_{j_R, \bar{\sigma}} a_{j_R, \sigma} \right) \\
    &+ \sum_{j_R} a_{j_R, \sigma} \left( \sum_{i_L} -t (e^{-J} - 1) n_{i_L, \bar{\sigma}} a_{i_L, \sigma}^{\dagger} \right)
\end{aligned}
\end{equation}
The above decomposition illustrates interactions that span across the boundary between the left and right blocks, with three operators acting within one block and one operator acting in the other. To reduce the bond dimension for this term, we define a complementary operator:
\begin{equation}
\hat{Q}_{ijkl} = \sum_{p_L, q_L, r_R} U a_{p_L, \uparrow}^\dagger a_{q_L, \downarrow} n_{r_R, \bar{\sigma}}.
\end{equation}
This complementary operator aggregates these three-body interactions across blocks to minimize the number of independent elements in the MPO representation. The bond dimension for $H_3$ can then be represented as:
\begin{equation}
M_{O,3} = 2\min\left(\frac{n_L^2(n_L-1)}{2}, n_R\right) + 2 \min\left(n_L, \frac{n_R^2(n_R-1)}{2}\right).
\end{equation}
Where the first term reflects the number of combinations for choosing three operators from one block with one operator from the other block. The second term considers pairwise interactions in a similar manner, ensuring a compact representation. The total bond dimension $M_O$ for the MPO representation of the transcorrelated Fermi-Hubbard Hamiltonian is the sum of the contributions from $H_1$, $H_2$, and $H_3$:
\begin{equation}
M_{O,\text{total}} = M_{O,1} + M_{O,2} + M_{O,3}.
\end{equation}
Assuming that the left and right blocks have equal sizes ($n_L = n_R = N/2$), we simplify the expression as:
\begin{equation}
M_{O,\text{max}} = 4 \left(\frac{N}{2}\right)^2 + 3 \left(\frac{N}{2}\right) + 2.
\end{equation}
In summary for the FH Hamiltonian, we have
\begin{itemize}
    \item $H_1$ represents the intra-block terms with bond dimension $M_{O,1} = 4$.
    \item $H_2$ represents two-body interactions between the blocks. A complementary operator technique is used to aggregate inter-block terms to reduce the bond dimension.
    \item $H_3$ represents three-body interactions between the blocks, which are more complex, and complementary operators are used to manage the number of independent combinations.
\end{itemize}
The indices in the complementary operators represent lattice sites in either the left or right block, and their role is to aggregate multiple interactions in order to minimize the bond dimension of the MPO representation. This detailed analysis allows for an efficient MPO representation of the transcorrelated Fermi-Hubbard model, keeping the bond dimension manageable even for the more complex terms introduced by the transcorrelated transformation. We previously computed the bond dimension contributions for both the Fermi-Hubbard Hamiltonian and the additional transcorrelated term. Since both contributions yield the same bond dimension, we can write the total bond dimension for the transcorrelated Fermi-Hubbard model as:
\begin{equation}
    M_{TC} = M_{\text{FH}} + M_{\text{Extra}}.
\end{equation}
Where $M_{\text{FH}}$ is the bond dimension from the Fermi-Hubbard model. $M_{\text{Extra}}$ is the bond dimension from the extra transcorrelated term. Since \(M_{\text{FH}} = M_{\text{Extra}}\), we have
\begin{equation}
    M_{TC} = 2 \times M_{\text{FH}}
\end{equation}
As we can see, these values show how the total bond dimension grows quadratically with the number of orbitals, but complementary operators help to manage this growth, ensuring that the computational complexity remains feasible for practical DMRG calculations. The introduction of complementary operators aggregates multiple terms, reducing the number of independent matrix elements required for the MPO representation, ultimately leading to an efficient and compact representation of the transcorrelated Fermi-Hubbard model. The fact that the transcorrelated terms contribute the same bond dimension as the original Fermi-Hubbard model is a positive outcome, indicating that the complexity of the MPO representation is not significantly increased. This is consistent with the purpose of the transcorrelated transformation, which aims to compactify the fermionic wavefunction, leading to a more efficient representation while keeping the computational requirements within manageable limits. The complementary operators used in the decomposition effectively group multiple interaction terms, helping control the growth of the bond dimension.
\begin{center}
Does the Bond Dimension Remain Manageable?
\par\end{center}
Despite adding the transcorrelated terms, the total bond dimension remains close to the non-transcorrelated case. This can be explained by several factors:

\subsection*{1. Types of Interactions Remain Local}
The additional transcorrelated terms still maintain the nearest-neighbor structure. This locality is crucial in ensuring that the MPO representation does not become overly complex. The hopping terms are modified by factors involving density operators, but they still act only between neighboring sites.

\subsection*{2. Complementary Operators}
The complementary operator technique effectively absorbs the changes introduced by the transcorrelated coefficients. These operators aggregate multiple similar interaction terms, thereby avoiding an explosion in the number of distinct elements that the MPO needs to store.

\subsection*{3. Compactification Effect}
The transcorrelated transformation has a natural role of compactifying the fermionic wavefunction, which is reflected in the MPO representation. Instead of introducing completely new kinds of operators or long-range interactions, the transformation modifies the strength of existing interactions, which allows us to represent the resulting MPO with similar computational complexity.

In summary, This result highlights the effectiveness of the transcorrelated approach in managing electron correlation effects without significantly increasing the MPO bond dimension. The transcorrelated Fermi-Hubbard model achieves a more compact representation of the fermionic wavefunction, which is an advantageous feature when it comes to practical DMRG calculations.

\section{\label{6} Conclusion and Outlook}

In this work, we introduced the EMTC-DMRG, a classical variational algorithm, which optimizes the ground-state wavefunction of the Transcorrelated Hamiltonian. Our algorithm is specifically designed for the  TI-DMRG problem, which operates outside the bounds of the variational principle due to the non-Hermitian nature of the TC-FH Hamiltonian. By leveraging a numerically exact optimization approach for the fermionic MPO, our method offers a powerful tool capable of surpassing established results under certain conditions, even when compared to more compact k-space formulations.

We demonstrated that in both weakly and strongly correlated regimes, our methodology achieved notable energy convergence while significantly reducing the computational resources required. Specifically, the initial bond dimension needed for our algorithm was substantially lower than those reported in the literature. Unlike competing methods that rely on k-space formulations, our approach operates in real space, addressing a key limitation of those alternatives. For more demanding structures, such as 6x6 lattices with higher filling, we suggest that our method could benefit from describing the target Hamiltonian in k-space.

Our results align with findings in the literature, such as those from FCIQMC that demonstrated that the ground states obtained via  TC Fermi-Hubbard Hamiltonians \cite{dobrautz, ref_11_section2_tc}, as well as TC ab initio DMRG methods \cite{TC_DMRG_Alavi_liao2023density, baiardi}, are efficiently represented by compact many-body wavefunctions thanks to transcorrelation approach. The successful application of the EMTC- DMRG algorithm to the 2D FH Hamiltonian suggests that this approach could be equally effective for electronic Hamiltonians. In particular, the Jastrow factor is conveniently expressed in real space, as the similarity transformed Hamiltonian includes up to three-body terms, with the primary term involving transcorrelated integrals.

A key innovation of our work lies in the use of an MPO formulation proposed by \cite{bipartite}, which proved highly efficient for handling ab initio Hamiltonians due to the absence of numerical errors through non-compression. In our work, we proved that this methodology is also  quite efficient when we are dealing with Hamiltonians  that contain three body terms. This makes our proposed algorithm exceptionally compact when combined with the transcorrelation approach. In future work, our aim is to explore initializing our algorithm with projected wavefunctions instead of random MPS initializations.

Projected wavefunction initialization involves starting with a trial wavefunction projected onto a lower-dimensional space, approximating the ground state more accurately and efficiently \cite{ref_11_pw, ref_13_pw, ref_14_pw, ref_17_pw, ref_18_pw, ref_19_pw, ref_41_pw, projectedwavefunction, projectedwavefunction_2}. By using a physically motivated trial wavefunction, the algorithm can achieve higher accuracy with fewer iterations, reducing computational time and resources. In contrast, random MPS initialization, the default in many libraries for classical variational algorithms, often faces challenges such as slower convergence, potential inaccuracies, and higher computational costs. While random MPS lacks the ability to capture the essential physics of the system effectively, projected wavefunctions offer significant advantages, particularly when prior knowledge of the system is available.

In summary, projected wavefunction methods are generally preferred over random MPS initialization due to their superior accuracy, efficiency, and reduced computational cost. However, the choice between these methods ultimately depends on the specific requirements and the knowledge available for the system under study.

Moving forward, we plan to apply our methodology to explore new frontiers in quantum thermodynamics, including the study of quantum phase transitions in 2D Fermi-Hubbard models. It would be interesting to explore simulations of 2D materials in the many-body regime based on massless Dirac fermions \cite{de_Moraes_2020}, for instance.
Another promising avenue involves extending this algorithm to investigate ground and excited states of molecular systems, possibly changing the Davison solver to the generalized version created to deal with TC Hamiltonians \cite{TC_DMRG_Alavi_liao2023density}. These applications highlight the versatility and potential of the EMTC DMRG algorithm to contribute significantly to several relevant research directions.

 \section*{Acknowledgments}

BGMA expresses gratitude to Akimasa Miyake for the opportunity to spend a year in his group, for providing a challenging and independent research direction, for encouraging intellectual risk-taking, and for his patience.

We extend our sincere thanks to Jiajun Ren for his invaluable guidance in using the Renormalizer library and for the insightful discussions that significantly contributed to this work. BGMA also acknowledges Marco A. Rodríguez-García for dedicating substantial time to running numerous simulations involving the iEMTC-DMRG and iTC-DMRG methods and for making her stay at CQuIC and in Albuquerque more enjoyable.

The authors thank Aron W. Cummings for his excellent feedback and constructive criticism, which improved the quality of this manuscript.

We also thank Sam McArdle and Werner Doubratz for their discussions. Additionally, we appreciate Hong-Hao Tu for providing excellent video and PDF lectures on tensor networks, as well as for valuable discussions on projected wavefunctions. 

We thank Guido Raos, Piero Macchi, Mosè Casalegno, and Alessandro Genoni for their guidance on electron correlation problems.
Finally, we express our gratitude to Chung-Yun Hsieh for carefully reviewing parts of this work, providing thoughtful feedback, and offering encouraging support.

This work was supported by the EAGER NSF CHE-2037832 grant (BGMA), CNPq grants 150963/2024-6 (BGMA), and 307626/2022-9 (AMSM).

\selectlanguage{english}%

\section{Appendix: Quantum Chemistry Methods \label{complemento_info}}

Electron correlation refers to the complex interactions between electrons in a many-electron system that are not fully captured by simple mean-field theories such as HF. Addressing electron correlation is essential for accurately describing the electronic structure and properties of atoms and molecules. Several advanced methods have been developed to tackle electron correlation, each with its own strengths and computational demands. For further information about these methods the reader can check these pedagogical references \cite{App1_Szabo1996, App2_Surjan1989, App3_Helgaker2000, App4_Jensen2007}.

\subsection{Approaches to Address Electron Correlation}

\subsubsection{Configuration Interaction (CI) Method}

The Configuration Interaction (CI) method expresses the electronic wave function as a linear combination of configuration state functions (CSFs), which are spin- and space-symmetry-adapted combinations of determinants. The CI coefficients are variationally determined by minimizing the energy expectation value with respect to the CSFs. Different levels of excitation (single, double, triple, etc.) relative to a reference CSF are included in the CI wave function. Large CI calculations involving thousands to billions of CSFs are routine, focusing on specific correlations while potentially excluding core orbitals to streamline computations.

\subsubsection{Perturbation Theory}

Perturbation theory expands the wave function in terms of a small parameter, typically the electron-electron interaction. It provides corrections to the HF energy by considering perturbations due to electron-electron interactions. Perturbative methods include Møller-Plesset perturbation theory (MP2, MP3, etc.) and coupled-cluster perturbation theory. These methods systematically improve upon the HF approximation by incorporating electron correlation effects.

\subsubsection{Coupled-Cluster (CC) Method}

The Coupled-Cluster (CC) method constructs an exponential ansatz for the wave function, systematically including both single and multiple excitations. Common variants like CCSD (singles and doubles) and CCSD(T) (singles, doubles, and perturbative triples) are widely used for their accuracy in capturing electron correlation effects.

\subsection{Post-HF Methods and Parametrization}

While HF theory, being the first successful \textit{ab initio} approach, posed significant challenges to computational chemists, its inherent limitations in addressing electron correlation necessitated the development of post-HF methods. These methods aim to treat the correlated motion of electrons more accurately than HF, which considers electron-electron interactions in a mean-field sense.

\subsubsection{Electron Correlation Methods}

The correlation energy, defined by Löwdin as the difference between the exact eigenvalue of the Hamiltonian and its HF approximation, represents the energy contribution from electron correlation:

\[
E_{corr} = E_{exact} - E_{HF}
\]
Since HF theory provides a well-defined energy that converges with an infinite basis set, the correlation energy accounts for the difference between this HF limit and the actual non-relativistic energy.

\subsection{Dynamic vs. Non-Dynamical Electron Correlation}

Electron correlation can be categorized into dynamic and non-dynamical (static) correlation:

\begin{itemize}
    \item \textbf{Dynamic Correlation:} Arises from the correlated motion of electrons due to their mutual repulsion, often addressed by methods like MP2 and multideterminant CI.
    \item \textbf{Non-Dynamical (Static) Correlation:} Occurs when a single-determinant reference is insufficient, as in systems with near-degenerate orbitals (e.g., singlet diradicals). This requires multiconfigurational methods like complete active space self-consistent field (CASSCF).
\end{itemize}

\subsubsection{The Møller-Plesset (MP) Approach}

The MP approach uses perturbation theory to iteratively improve upon the HF energy, with levels MP1, MP2, MP3, and so on, each providing progressively better approximations of electron correlation effects. MP2, for example, adds a correction term $E^{(2)}$ to the HF energy to account for electron correlation.

\[
E_{MP2} = E_{total}^{HF} + E^{(2)}
\]

\subsubsection{Configuration Interaction (CI)}

CI wavefunctions are constructed by promoting electrons from occupied to unoccupied orbitals, forming a multi-determinant wavefunction. CI calculations can be highly accurate but are computationally intensive. Variants like CI single excitations, CI singles and doubles, and full CI (all possible excitations) are used depending on the desired accuracy and computational resources.

\subsubsection{Multi-Configurational Self-Consistent Field}

MCSCF methods optimize orbitals for a multiple-determinant wavefunction, providing accurate correlation energies with fewer configurations compared to CI. The CASSCF method, which includes all combinations of active space orbitals, is particularly effective for capturing valence region correlations.

\subsubsection{Coupled-Cluster (CC)}

CC methods, such as CCSD and CCSD(T), use an exponential ansatz for the wavefunction, combining multiple determinants to capture electron correlation more comprehensively than CI.

\subsection{Strengths and Weaknesses of \textit{Ab Initio} Methods}

\subsubsection{Strengths}

\begin{itemize}
    \item \textit{Ab initio} calculations are based on the Schrödinger equation, providing accurate and reliable results without empirical adjustments.
    \item These methods can be applied to various molecular species, including transition states and non-stationary points.
    \item The accuracy of \textit{ab initio} methods can be systematically improved by increasing the basis set size and using higher-level post-HF methods.
\end{itemize}

\subsubsection{Weaknesses}

\begin{itemize}
    \item \textit{Ab initio} calculations are computationally intensive and require significant resources.
    \item The need for extensive computational power and memory can limit the applicability of \textit{ab initio} methods for large systems.
\end{itemize}

\textit{Ab initio} methods form the foundation for solving the Schrödinger equation in quantum chemistry. While the HF method provides a starting point, post-HF methods like MP, CI, and CC are essential for accurately treating electron correlation. These methods enable the calculation of molecular geometries, energies, vibrational frequencies, and various spectroscopic properties, contributing significantly to our understanding of molecular systems.

\end{document}